\documentclass[aps,superscriptaddress,twocolumn]{revtex4-1}

\usepackage[pdftex]{graphicx}
\usepackage{dcolumn}
\usepackage{bm}
\usepackage{color}
\usepackage{amsmath}
\usepackage{nicefrac}
\usepackage{amssymb}

\usepackage{braket}
\usepackage{physics}
\usepackage{bbold}
\usepackage[pdftex,colorlinks=true,linkcolor=blue,citecolor=blue,filecolor=blue]{hyperref}

\newcommand{\ff}[1]{{\boldsymbol #1}}
\newcommand{\ca}[1]{{\cal #1}}
\newcommand{\bi}{\begin{itemize}}
\newcommand{\ei}{\end{itemize}}
\newcommand{\be}{\begin{equation}}
\newcommand{\ee}{\end{equation}}
\newcommand{\ba}{\begin{eqnarray}}
\newcommand{\ea}{\end{eqnarray}}

\newcommand{\refeq}[1]{Eq.\ (\ref{eq:#1})}
\newcommand{\labeq}[1]{\label{eq:#1}}

\begin{document} 
  
\title{Emergent Non-Abelian Gauge Theory in Coupled Spin-Electron Dynamics}

\author{Nicolas Lenzing}

\affiliation{I. Institute of Theoretical Physics, Department of Physics, University of Hamburg, Notkestra{\ss}e 9-11, 22607 Hamburg, Germany}

\author{Alexander I. Lichtenstein}

\affiliation{I. Institute of Theoretical Physics, Department of Physics, University of Hamburg, Notkestra{\ss}e 9-11, 22607 Hamburg, Germany}

\affiliation{The Hamburg Centre for Ultrafast Imaging, Luruper Chaussee 149, 22761 Hamburg, Germany}

\author{Michael Potthoff}

\affiliation{I. Institute of Theoretical Physics, Department of Physics, University of Hamburg, Notkestra{\ss}e 9-11, 22607 Hamburg, Germany}

\affiliation{The Hamburg Centre for Ultrafast Imaging, Luruper Chaussee 149, 22761 Hamburg, Germany}

\begin{abstract}
A clear separation of the time scales governing the dynamics of ``slow'' and ``fast'' degrees of freedom often serves as a prerequisite for the emergence of an independent low-energy theory.
Here, we consider (slow) classical spins exchange coupled to a tight-binding system of (fast) conduction electrons. 
The effective equations of motion are derived under the constraint that the quantum state of the electron system at any instant of time $t$ lies in the $n$-dimensional low-energy subspace for the corresponding spin configuration at $t$.
The effective low-energy theory unfolds itself straightforwardly and takes the form of a non-abelian gauge theory with the gauge freedom given by the arbitrariness of the basis spanning the instantaneous low-energy sector.
The holonomic constraint generates a gauge covariant spin-Berry curvature tensor in the equations of motion for the classical spins. 
In the non-abelian theory for $n>1$, opposed to the $n=1$ adiabatic spin dynamics theory, the spin-Berry curvature is generically nonzero, even for time-reversal symmetric systems.
Its expectation value with the representation of the electron state is gauge invariant and gives rise to an additional {\em geometrical} spin torque.
Besides anomalous precession, the $n\ge 2$ theory also captures the spin nutational motion, which is usually considered as a retardation effect.
This is demonstrated by proof-of-principle numerical calculations for a minimal model with a single classical spin.
Already for $n=2$ and in parameter regimes where the $n=1$ adiabatic theory breaks down, we find good agreement with results obtained from the full (unconstrained) theory.
\end{abstract} 

\maketitle 

\section{Introduction}
\label{sec:intro}

Classical spin models \cite{Now07,BMS09} are a highly useful and widely employed tool to understand the non-equilibrium dynamics of magnetic materials.
At the expense of disregarding the quantum nature of the magnetic moments and related phenomena, such as the Kondo effect \cite{Kon64,Hew93}, they provide a numerically tractable framework for spin dynamics on an atomistic length scale
\cite{TKS08, SHNE08, FI11, EFC+14}.
Typically, classical spin models may comprise a short-range isotropic Heisenberg-type exchange, various anisotropic couplings and long-range, e.g., dipole interactions. 
The classical equations of motion are usually supplemented by Gilbert-damping terms to account for dissipation effects.

Spin-only models can actually be seen as effective low-energy theories emerging from a more fundamental level of modelling, where the local magnetic moments (classical spins $\ff S_{i}$) at sites $i$ of a lattice are coupled to the local spins $\ff s_{i}$ of a system of conduction electrons via a local exchange coupling $J$. 
Such quantum-classical spin-electron hybrid models are necessary to explain various phenomena, including 
indirect spin exchange interactions, like the Ruderman-Kittel-Kasuya-Yoshida (RKKY) interaction \cite{rkky},
Gilbert spin damping due to coupling to electronic degrees of freedom \cite{llg},
spin inertia effects (nutation) \cite{But06,WC12},
and other more strongly retarded effective spin-spin interactions mediated by the conduction-electron system.

The standard formal approach \cite{ON06,UMS12,BNF12,SP15,BN19} that achieves the derivation of the effective spin-only theory is based on the (usually realistic) assumption that the local exchange coupling $J$ is weak as compared to the typical energy scales of the electron system. 
Consider the $s$-$d$ model \cite{vz} with Hamiltonian $H=H_{\rm el.} + J \sum_{i} \ff s_{i} \ff S_{i}$ as a prototype. 
The torque on the classical spin at site $i$ is given by $J \langle \ff s_{i} \rangle_{t} \times \ff S_{i}$, where the expectation value of the local electron spin $\ff s_{i}$ at site $i$ is obtained from the many-body state $| \Psi(t) \rangle$ of the electron system (Hamiltonian $H_{\rm el.}$) at time $t$.
Since the electron state itself must be computed in the presence of the local exchange interaction term $\propto J$ for the (time-dependent) classical spin configuration $\{ \ff S \}$, there is a retarded mutual effective interaction emerging. 
This is uncovered, for example, by linear-response theory, i.e., by lowest-order time-dependent perturbation theory in $J$. 
This leads to an integro-differential equation of motion for $\ff S_{i}$, 
\be
\dot{\ff S}_{i}(t) = J^{2} \sum_{i'}
\int_{0}^{t} dt^\prime \underline{\chi}_{ii'}(t - t^\prime) \boldsymbol{S}_{i'}(t^\prime) \times \ff S_{i}(t) 
\label{eq:integro}
\ee
which involves the retarded magnetic susceptibility tensor with elements ${\chi}_{ii'}^{(\alpha\alpha')}(t-t')$ of the electron ground state as the integral kernel ($\alpha,\alpha'=x,y,z$).
The resulting spin dynamics is non-conservative, as Eq.\ (\ref{eq:integro}) describes an open quantum system and is known from Redfield theory \cite{BP02}.

Assuming that $\underline{\chi}_{ii'}(t - t^\prime)$ is strongly peaked at $t' = t$, 
we can replace $\ff S_{i'}(t')$ by the first few terms in its Taylor expansion around $t'=t$, i.e., 
$\ff S_{i'}(t') \approx \ff S_{i'}(t) + \dot{\ff S}_{i'}(t) (t'-t) + \ddot{\ff S}_{i'}(t) (t'-t)^{2}/2$.
Keeping the first term on the right-hand side only and extending the integration over $t'$ to infinity, one obtains an effective Hamiltonian equation of motion for the spins $\ff S_{i}$, which involves the instantaneous spin-spin interaction mediated by the RKKY coupling $J_{ii'}^{\rm (RKKY)} = J^{2} \chi_{ii'}(\omega=0)$. 
Including the second term in addition, gives rise to a (non-local) Gilbert damping tensor $\underline{\alpha}_{ii'} = - i J^{2} \partial_{\omega} \underline{\chi}_{ii'}(\omega)|_{\omega=0}$, while the third term leads to spin-inertia effects, i.e., additional nutation of the spins.
This derivation has been put forward in Refs.\ \cite{BNF12,SP15} and can be employed in the context of strongly correlated electron models \cite{SRP16a} or, when combined with band-structure theory, for an {\em ab initio} computation of the Gilbert damping \cite{AKvSH95,KK02,CG03,EMKK11}.
Nutation effects, as have been discussed in Refs.\ \cite{FSI11,KT15,SRP16b}, for example, find a natural explanation in the same framework set by Eq.\ (\ref{eq:integro}).
Furthermore, at least in principle, systematic extensions of the resulting low-energy spin-only theory can be achieved by taking into account terms of higher order in the expansion. 
One may also drop the approximation on the $t'$-integration range. 
This leads to a time-dependent RKKY coupling $J^{(RKKY)}_{ii'}(t)$ and a time-dependent Gilbert damping $\underline{\alpha}_{ii'}(t)$, as has been mentioned in Ref.\ \cite{SP15,BN19}.

The above-sketched standard theory misses, however, an important effect pointed out recently \cite{SP17}: 
The slow dynamics of the classical spins results in a non-trivial Berry curvature of the electronic quantum system as is well known since long \cite{Ber84,XCN10}.
Quite generally, however, this Berry curvature in turn has a feedback on the classical spin dynamics \cite{WZ88, NK98, BMK+03, NWK+99, SP17}.
Namely, there is a geometrical spin torque which comes with the same prefactor $J^{2}$ as the RKKY coupling and the Gilbert damping. This torque can give rise to unconventional spin dynamics as has been demonstrated \cite{SP17,BN20} not only for a quantum-classical system as is considered here as well, but also for slow classical spins locally exchange coupled to a system of fast classical spins \cite{EMP20,MP21} and even for the dynamics of a quantum spin in a Kondo model \cite{SP17}.

This geometrical spin torque emerges in an effective low-energy spin-only theory that is derived by starting from full theory of classical spins coupled to conduction electrons by imposing the constraint that, at any instant of time $t$, the electron system is in its ground state, i.e., $|\Psi(t) \rangle = | \Psi_{0}(\{\ff S(t)\} ) \rangle$, for the spin configuration $\{\ff S(t)\}$ at time $t$.
This is analogous to molecular dynamics approaches \cite{MH00,BMK+03,ZW06} where the slow nuclear coordinates are treated classically.
If the exchange coupling $J$ is weak, the classical spin dynamics is slow compared to typical energy scales of the electron systems.
The adiabatic spin dynamics (ASD) thus addresses the same parameter regime as the standard perturbative linear-response approach discussed above. 

With the present paper we explore a systematic extension of the ASD by relaxing the adiabatic constraint. 
The impact of electronic low-energy excitations from the instantaneous ground state $| \Psi_{0}(\{\ff S(t)\}) \rangle$ on the classical spin dynamics can be taken into account by imposing, as a weaker constraint, that the electron state $| \Psi(t) \rangle$ be at time $t$ in the subspace of the Fock space spanned by the first $n>1$ eigenstates of the Hamiltonian for the spin configuration $\{\ff S(t)\}$ at $t$.
This beyond-adiabatic constraint leads to a non-abelian Berry connection and curvature \cite{XCN10, WZ84}. 
Here, we will work out the general formalism of the non-abelian gauge theory that emerges as the effective low-energy theory. 
The formally correct incorporation of the constraint is achieved within conventional Lagrange formalism. 
A simple toy model will be considered and solved numerically to study the effect of the geometric torque on the classical spin dynamics in the non-abelian case.
We discuss the anomalies in the precessional spin dynamics and demonstrate that spin nutation arises naturally in our framework. 
The previously developed ASD represents the $n=1$ limit of our non-abelian spin-dynamics (NA-SD) theory. 
In the ASD for a single classical spin, the presence of an anomalous precession frequency has been found \cite{SP17} for an odd number of conduction electrons only, while the full solution of the coupled equations of motion for spin and electron dynamics yields an anomalous frequency for both, odd and even electron numbers. 
In the broader framework of NA-SD we can resolve this open issue. 

The paper is organized as follows: 
The next section \ref{sec:eff} presents the general Hamiltonian and Lagrangian formulation of the theory. 
The equations of motion of the non-abelian gauge theory in the instantaneous low-energy sector are worked out in Sec.\ \ref{sec:eom}, and various formal aspects of the theory are discussed in Sec.\ \ref{sec:dis}.
Sections\ \ref{sec:tr} and \ref{sec:gauge} are particularly devoted to a discussion of the impact of time-reversal symmetry and of gauge transformations, respectively. 
A minimal model, suitable for proof-of-principle studies, is introduced in Sec.\ \ref{sec:mod}. 
In Sec.\ \ref{sec:res} we present and discuss the results of numerical calculations. 
Conclusions are given in Sec.\ \ref{sec:con}.

\section{General Theory}
\label{sec:eff}

Geometric forces or torques originate in the adiabatic limit of hybrid systems consisting of quantum degrees of freedom interacting with classical degrees of freedom.
Here, we consider a quantum lattice model of $N$ conduction electrons interacting with $M$ classical ``spins'' $\ff S_{m}$ of unit length $|\ff S_{m}|=1$. 
The system dynamics is governed by a quantum-classical Hamiltonian of the form 
\be
\hat{H} (\{\boldsymbol{S}\})
= 
\hat{H}_{\text{qu}}
+ {H}_{\text{cl}}(\{\boldsymbol{S}\}) 
+ \hat{H}_{\rm int}(\{\boldsymbol{S}\})
\: .
\labeq{ham}
\ee
The quantum Hamiltonian $\hat{H}_{\text{qu}}$ is constructed in terms of fermion creation and annihilation operators $c^{\dagger}_{\ff r \sigma}$ and $c_{\ff r \sigma}$, where $\ff r$ refers to the sites of the lattice and $\sigma = \uparrow, \downarrow$ is the spin projection. 
Additional orbital degrees of freedom may be considered as well. 
The formulation of the theory is largely independent of $\hat{H}_{\text{qu}}$ but requires a well-defined local quantum spin $\ff s_{\ff r}$ at lattice site $\ff r$:
\be
  \ff s_{\ff r} = \frac12 \sum_{\sigma\sigma'} c^{\dagger}_{\ff r \sigma} \ff \sigma_{\sigma \sigma'} c_{\ff r \sigma'} 
  \: .
\ee 
Here, $\ff \sigma$ is the vector of $2\times 2$ Pauli matrices (and $\hbar \equiv 1$).

The dynamics of the subsystem of $M$ classical spins $\{ \ff S \} \equiv \{ \ff S_{1}, ..., \ff S_{M} \}$ derives from a classical Hamilton function ${H}_{\text{cl}}(\{\boldsymbol{S}\})$ and may comprise an external magnetic field and isotropic or anisotropic spin exchange couplings. 
The third term in \refeq{ham} represents a quantum-classical interaction term. 
Here, we choose an isotropic local exchange interaction 
\be
  \hat{H}_{\text{int}}(\{\boldsymbol{S}\}) = J \sum_{m=1}^{M} \boldsymbol{S}_{m} \boldsymbol{s}_{\ff r_{m}}
  \: , 
\labeq{hint}  
\ee
between the $m$-th classical spin $\ff S_{m}$ and the local spin $\ff s_{\ff r_{m}}$ of the conduction-electron system at the site $\ff r_{m}$.
The coupling strength is $J>0$.
The theory is developed for an arbitrary number of classical spins $M$, but we will later focus on a single-classical-spin Kondo model ($M=1$) for the sake of simplicity.

If the classical spins $\{\ff S\}$ were replaced by quantum spins, \refeq{ham} would represent the Hamiltonian of the multi-impurity or lattice Kondo model. 
With the classical-spin approximation we disregard typical correlation effects, such as Kondo screening and heavy-fermion behavior, and hence we are essentially working on a mean-field-type level.
The approximation may be justified in cases where there are well-formed spin moments which are stable on time scales exceeding all remaining time scales of the problem, e.g., in cases, where the Kondo effect is suppressed by magnetism or in case of quantum spins with large spin quantum numbers. 
An example has been given in Ref.\ \cite{SRP16a}, where anomalous quantum-classical dynamics due to a geometrical torque has also been found in the corresponding full quantum system.
A consistent theory for a system that is entirely quantum with a least two largely different time scales has yet to be developed. 
This means that the presence of slow classical degrees of freedom is necessarily required for the very concept of geometrical forces and torques. 
The classical degrees of freedom are required to define the smooth manifold onto which the quantum dynamics is restricted in the adiabatic limit. 

A pure state of the quantum-classical hybrid system at time $t$ is specified by a Hilbert-space vector $| \Psi(t) \rangle$ and by the classical spin configuration $\{ \ff S(t) \}$, see Refs.\ \cite{Hes85, Hal08, Elz12} for a general discussion of hybrid dynamics.
The trajectory of the system state is obtained as the solution of a system of coupled ordinary differential equations. 
These consist of the Schr\"odinger equation, involving the quantum Hamiltonian and the interaction term, which depends on the classical-spin configuration,
\be
i \partial_{t} \ket{\Psi(t)} = [\hat{H}_{\text{qu}} + \hat{H}_{\rm int}(\{\boldsymbol{S}(t)\}) ] \ket{\Psi(t)} 
\: ,
\labeq{eom1}
\ee
and the Hamilton equations of motion for the classical-spin configuration, involving the classical Hamilton function and the expectation value of the interaction term in the quantum state $| \Psi(t) \rangle$:
\be
\dot {\ff S}_{m}(t) = \Big\{ \ff S_{m}(t) , {H}_{\text{cl}}(\{\boldsymbol{S}(t)\}) + \langle \hat{H}_{\rm int}(\{\boldsymbol{S}(t)\}) \rangle \Big\}_{S}
\: .
\labeq{eom2}
\ee
Here, the dot denotes the time derivative, and $\{ \cdot , \cdot \}_{S}$ is the Poisson bracket. 
In case of spin systems, the latter is defined for two arbitrary functions $A(\{\boldsymbol{S}\})$ and $B(\{\boldsymbol{S}\})$ as \cite{BK90}
\be
\{ A, B \}_{S} = \sum_{m} 
\frac{\partial A}{\partial \ff S_{m}}
\times
\frac{\partial B}{\partial \ff S_{m}}
\cdot
\ff S_{m} \: .
\labeq{poisson}
\ee

The coupled equations of motion, \refeq{eom1} and \refeq{eom2}, are generated as Euler-Lagrange equations by requiring stationarity of an action functional $\ca S = \int L dt$ with the Lagrangian $L = L(\{\boldsymbol{S}\}, \{\dot{\boldsymbol{S}}\},\ket{\Psi},\dot{\ket{\Psi}},\bra{\Psi},\dot{\bra{\Psi}})$: 
\be
L = \sum_m\boldsymbol{A}(\boldsymbol{S}_m)\dot{\boldsymbol{S}}_m + \bra{\Psi(t)}i\partial_{t} - \hat{H}\ket{\Psi(t)}
\: . 
\labeq{lag}
\ee
Here, $\ff A(\ff S)$ is a function satisfying $\nabla \times \ff A(\ff S) = - \ff S / S^{3}$, which can thus be interpreted as the vector potential of a unit magnetic monopole located at $\ff S = 0$. We have 
\be
  \ff A(\ff S) = - \frac{1}{S^{2}} \frac{\ff e \times \ff S}{1+\ff e \ff S / S} \; , 
\labeq{mono}
\ee
with a unit vector $\ff e$.
In the standard gauge \cite{Dir31} this is chosen as $\ff e = \ff e_{z}$.
In this gauge, another representation is $\ff A(\ff S)=- (1/S) \tan (\vartheta/2) \ff e_{\rm \varphi}$, using 
spherical coordinates $(S, \vartheta, \varphi)$.
For details of deriving \refeq{eom1} and \refeq{eom2} from $\delta \ca S = 0$, see Ref.\ \cite{EMP20} (supplemental material).

We will address the parameter regime of the Hamiltonian, where the system dynamics is characterized by two strongly different time scales, a slow spin dynamics and a fast dynamics of the electron state, which almost instantaneously follows the motion of the spins. 
In the extreme adiabatic limit, the quantum many-body state  $|\Psi(t) \rangle$ of the electron system at time $t$ is given by the ground state, $| \Psi_{0}(\{\ff S(t)\} ) \rangle$ of $\hat{H}_{\text{qu}} + \hat{H}_{\text{int}}(\{\boldsymbol{S}(t)\})$, for the spin configuration $\{\ff S(t)\}$ at time $t$.
When approaching the adiabatic limit in parameter space, the fast electron dynamics will be more and more constrained to the ground 
manifold $\{ | \Psi_{0}(\{\ff S(t)\} ) \rangle \}$.
Adiabatic spin-dynamics (ASD) theory \cite{SP17,EMP20,MP21} assumes that the dynamics is {\em perfectly} constrained to the ground-state manifold and employs
\be
  |\Psi(t) \rangle = | \Psi_{0}(\{\ff S(t)\} ) \rangle
\labeq{adiab}  
\ee
as a holonomic constraint to completely eliminate the electron degrees of freedom from the Lagrangian \refeq{lag}. 
In this way, one arrives at a spin-only effective Lagrangian $L_{\rm eff}(\{ \ff S \}, \{ \dot{\ff S }\})$, and the resulting effective equations of motion include the geometrical spin torque as an holonomy effect \cite{SP17}.
The unconventional spin dynamics originating from the corresponding geometrical spin torque is missed by other approaches, such as the standard linear-response approach to a spin-only theory that has been discussed in the introduction. 
On the other hand, retardation effects, e.g., nutational motion, are excluded within ASD by the very construction. 

The validity of the basic assumption, \refeq{adiab}, strongly depends on the specific system considered and on the considered parameter range. 
Even for gapped systems, however, the strict adiabatic approximation is never perfectly satisfied, and the true slow spin dynamics will be affected to some degree by admixtures from (low-energy) excited electron states. 
As a systematic generalization of ASD, we therefore propose to relax the constraint \refeq{adiab} and to replace it by the weaker constraint
\be
  \ket{\Psi(t)} = \sum_{i=0}^{n-1} \alpha_{i}(t) \ket{\Psi_{i}(\{\boldsymbol{S}(t)\})} \: .
\labeq{const}  
\ee
Here, $\ket{\Psi_{i}(\{\boldsymbol{S}(t)\})}$ is the $i$-th excited state of $\hat{H}_{\text{qu}} + \hat{H}_{\text{int}}(\{\boldsymbol{S}(t)\})$, i.e., we assume that at any instant of time $t$ the conduction-electron state $\ket{\Psi(t)}$ is contained in the low-energy subspace ${\cal E}_{n}(\{ \ff S\})$ spanned by the instantaneous ground state and the lowest $n-1$ instantaneous eigenstates for the spin configuration $\{\ff S\} = \{\ff S(t)\}$ at time $t$.
Choosing a fixed orthonormal basis 
\be
\{ \ket{\Psi_{i}(\{\boldsymbol{S}\})} \, | \, {i=0,...,n-1}\} 
\labeq{basis}
\ee
of ${\cal E}_{n}(\{\ff S\})$ for any spin configuration, the electron state at time $t$ is fully specified by the set of expansion coefficients $\{\alpha(t)\} \equiv \{ \alpha_{0}(t), ..., \alpha_{n-1}(t) \}$ via \refeq{const}.

For $n=1$, we recover conventional ASD, and thus obtain a true spin-only theory.
For small $n>1$, the effective Lagrangian is obtained from \refeq{lag} by substituting $\ket{\Psi}$, $\partial_{t} |{\Psi} \rangle$, $\bra{\Psi}$, $\partial_{t} \langle{\Psi}|$ using \refeq{const}.
It thereby becomes a function of $\{\ff S \}$ and $\{\dot{\ff S}\}$ and furthermore a function of the set of expansion coefficients $\{\alpha\}$, i.e., we get $L_{\text{eff}} = L_{\text{eff}}(\{\boldsymbol{S}\}, \{\dot{\boldsymbol{S}}\},\{\alpha\},\{\alpha^{\ast}\},\{\dot{\alpha}\}),\{\dot{\alpha}^{\ast}\})$.
Hence, besides the spin degrees of freedom, the resulting low-energy theory contains a few electronic degrees of freedom as well. 

We also define the eigenenergies $E_{i} = E_i (\{\ff S \})$ of $\hat{H}_{\text{qu}} + \hat{H}_{\text{int}}(\{\boldsymbol{S}(t)\})$ corresponding to the basis states $|\Psi_{i}(\{\boldsymbol{S}\}) \rangle$.
$E_i (\{\ff S \})$ is the analog of the $i$-th potential-energy (Born-Oppenheimer) surface known from molecular-dynamics theory  \cite{MH00,BMK+03}.
The spin configuration $\{\ff S\}$ takes the role of the configuration of atomic nuclei.
Note that the strict adiabatic approximation, \refeq{adiab}, becomes invalid, if the trajectory of the spin configuration $\{ \ff S(t) \}$ passes a configuration $\{ \ff S_{\rm cr} \}$, at which there is a crossing of the ground state with the first excited state, i.e., $E_{0}(\{ \ff S_{\rm cr} \}) = E_{1}(\{ \ff S_{\rm cr} \})$, since this is in conflict with the adiabatic theorem \cite{Kat50,AE99,Com09}. 

For $n>1$, the relaxed condition (\ref{eq:const}) corresponds to a generalized adiabatic theorem, see Ref.\ \cite{Kat50}, stating that the condition is respected, if the low-energy sector ${\cal E}_{n}(\{\ff S\})$ and its orthogonal complement (the ``high-energy sector'') remain gapped for all $\{ \ff S(t) \})$ and, of course, if the electron dynamics is sufficiently slow. 
In other words, for a given $n$, NA-SD applies if there is no crossing $E_{n-1}(\{\ff S_{\rm cr} \}) = E_{n}(\{ \ff S_{\rm cr} \})$, while crossings of states {\em within} the low-energy sector are irrelevant.
One should note, however, that a crossing of two states belonging to the low- and the high-energy sector, respectively, is in fact unproblematic, if the expansion coefficient $\alpha_{n-1}(t)=0$ for all $t$, since in this case the $n-1$-th excited eigenstate would not contribute to $| \Psi(t) \rangle$ anyway.
This argument can be extended to $k < n - 1$, as long as there are crossings between ``unoccupied'' states with $\alpha_{i}(t)=0$ and $\alpha_{j}(t) = 0$ for $k \le i,j \le n$ only. 
We conclude that the relaxed condition (\ref{eq:const}) for $n>1$ also implies a less severe, relaxed approximation.

\begin{widetext}

\section{Effective equations of motion}
\label{sec:eom}

The effective Lagrangian that is obtained by using the constraint \refeq{const} to eliminate $\ket{\Psi(t)}$ from the original Lagrangian \refeq{lag}, is given by: 
\be
  L_{\text{eff}} = L_{\text{eff}}(\{\boldsymbol{S}\}, \{\dot{\boldsymbol{S}}\},\{\alpha\},\{\alpha^{\ast}\},\{\dot{\alpha}\},\{\dot{\alpha}^{\ast}\})
  = 
  \sum_m\boldsymbol{A}_m(\boldsymbol{S}_m) \dot{\boldsymbol{S}}_m+i\sum_{ij}\alpha_{i}^{\ast}\bra{\Psi_{i}}\partial_{t}(\alpha_{j}  
  \ket{\Psi_{j}})-\sum_{ij}\alpha_{i}^{\ast}\alpha_{j}\bra{\Psi_{i}}\hat{H}\ket{\Psi_{j}} \: ,
\label{eq:leff}  
\ee
where $\ket{\Psi_{i}} = \ket{\Psi_{i}(\{\boldsymbol{S}_m\})}$, and where the $\{\dot{\ff S}\}$-dependence, besides the first term, is due to $\bra{\Psi_{i}}\partial_{t}\ket{\Psi_{j}} = \sum_m \bra{\Psi_{i}}\partial_{\ff S_{m}} \ket{\Psi_{j}} \dot{\ff S}_{m}$. 
The Euler-Lagrange equation $\partial_{t} (\partial L_{\text{eff}} / \partial \dot{\alpha}^{\ast}_{i}) - \partial L_{\text{eff}} / \partial\alpha^{\ast}_{i} = 0$ for the ``wave function'' $\alpha_{i}$ is straightforwardly obtained as:
\be
  i \partial_{t} \alpha_{i} 
  = 
  \sum_{j} \bra{\Psi_{i}} ( \hat{H}_{\text{qu}} + \hat{H}_{\text{int}} )\ket{\Psi_{j}} \alpha_{j} 
  - i \sum_{m}\sum_{j} \alpha_{j} \bra{\Psi_{i}} \partial_{\ff S_{m}} \ket{\Psi_{j}} \dot{\ff S}_{m}
\: .
\labeq{wf}
\ee
The complex conjugate of this equation is just the equation of motion that is obtained for $\alpha_{i}^{\ast}$.

Note that the second term involves the non-abelian spin-Berry connection $\underline{\ff C}_{m} = \underline{\ff C}_{m} (\{ \ff S \})$. 
Opposed to the (abelian) spin-Berry connection $\ff C_{m} = i \langle \Psi_{0} | \partial_{\ff S_{m}} | \Psi_{0} \rangle$ of the (abelian) ASD, this is, for each $m$, a matrix-valued vector with elements:
\be
\ff C^{(ij)}_{m} = i \bra{\Psi_{i}} \partial_{\ff S_{m}} \ket{\Psi_{j}} 
=
i \sum_{\gamma} \bra{\Psi_{i}} \partial_{S_{m\gamma}} \ket{\Psi_{j}}  \ff e_{\gamma}
=
\sum_{\gamma} C^{(ij)}_{m\gamma}\ff e_{\gamma}
\: .
\label{eq:con}
\ee
The matrix dimension is given by the dimension of the low-energy subspace $n=\dim {\cal E}_{n}(\{\ff S\})$.
It is easy to see that this is a real quantity.
Its transformation behavior under gauge transformations will be discussed in Sec.\ \ref{sec:gauge}.

We proceed by deriving the second set of equations of motion from the effective Lagrangian
$\partial_{t} (\partial L_{\rm eff} / \partial \dot{\ff S}_m) - \partial L_{\rm eff} / \partial \ff S_m = 0$.
With \refeq{leff} we straightforwardly find:
\be
\frac{\partial L_{\text{eff}}}{\partial \boldsymbol{S}_m} 
= 
\frac{\partial}{\partial \ff S_{m}} (\ff A_{m} \dot{\ff S}_{m})
+ 
i \sum_k\sum_{ij} \alpha^\ast_i \alpha_j \frac{\partial}{\partial \ff S_{m}} (\bra{\Psi_{i}} \partial_{\ff S_{k}} \ket{\Psi_{j}} \dot{\ff S}_{k} )
- 
\sum_{ij} \alpha^\ast_i \alpha_j \frac{\partial}{\partial \ff S_{m}} (\bra{\Psi_{i}} \hat{H} \ket{\Psi_{j}}) 
\: , 
\label{eq:lags}
\ee
and with $
\partial L_{\text{eff}} / \partial \dot{\boldsymbol{S}}_m
= 
\ff A_{m} 
+ 
i \sum_{ij} \alpha^\ast_i \alpha_j \bra{\Psi_{i}} \partial_{\ff S_{m}} \ket{\Psi_{j}} 
$, 
\be
\frac{\text{d}}{\text{d} t} \frac{\partial L_{\text{eff}}}{\partial \dot{\boldsymbol{S}}_m} 
= \sum_{\gamma} \frac{\partial \ff A_{m}}{\partial S_{m\gamma}} \dot{S}_{m\gamma} 
+ 
i \sum_{ij}  [(\partial_t \alpha^\ast_i)\alpha_j + \alpha^\ast_i (\partial_t \alpha_j)] \bra{\Psi_{i}} \partial_{\ff S_{m}} \ket{\Psi_{j}} 
+ 
i \sum_k \sum_{ij}  \alpha^\ast_i \alpha_j (\dot{\ff S}_{k} \partial_{\ff S_{k}}) (\bra{\Psi_{i}} \partial_{\ff S_{m}} \ket{\Psi_{j}} )
\: .
\labeq{lagsd}
\ee
Both, \refeq{lags} and \refeq{lagsd} involve the spin-Berry connection. 
The third term in \refeq{lags} can be rewritten using the identity
\be
\frac{\partial}{\partial \ff S_{m}} (\bra{\Psi_{i}} \hat{H} \ket{\Psi_{j}}) = \bra{\Psi_{i}} (\partial_{\ff S_{m}} \hat{H}) \ket{\Psi_{j}} - (E_j - E_i) \bra{\Psi_{i}} \partial_{\ff S_{m}} \ket{\Psi_{j}}
\: , 
\ee
and for the second term in \refeq{lagsd} it is convenient to get rid of the time derivatives by using
\be
i [(\partial_t \alpha^\ast_i)\alpha_j + \alpha^\ast_i (\partial_t \alpha_j)] 
= 
i \sum_{k} \sum_{l} \left[ \alpha^\ast_l \alpha_j \bra{\Psi_{l}} \partial_{\ff S_{k}} \ket{\Psi_{i}} 
-
\alpha^\ast_i \alpha_l \bra{\Psi_{j}} \partial_{\ff S_{k}} \ket{\Psi_{l}} \right] 
\dot{\ff S}_{k} 
+ 
\alpha^\ast_i \alpha_j (E_j - E_i)
\: , 
\ee
which directly follows from the equation of motion for the wave functions \refeq{wf}.
Therewith, we arrive at
\begin{align}
0 
&= 
\frac{\text{d}}{\text{d} t} \frac{\partial L_{\text{eff}}}{\partial \dot{\boldsymbol{S}}_m} - \frac{\partial L_{\text{eff}}}{\partial \boldsymbol{S}_m} \nonumber\\
&= 
\sum_{\beta\gamma} \left( \frac{\partial A_{m\beta}}{\partial S_{m\gamma}} - \frac{\partial A_{m\gamma}}{\partial S_{m\beta}} \right) \dot{S}_{m\gamma} \hat{e}_\beta 
+  
\sum_{ij} \alpha^\ast_i \alpha_j \bra{\Psi_{i}} (\partial_{\ff S_{m}} \hat{H}) \ket{\Psi_{j}} 
\nonumber\\
&+ 
i \sum_{k} \sum_{ij} \sum_{\gamma} \alpha^\ast_i \alpha_j 
\left[ \partial_{S_{k\gamma}} (\bra{\Psi_{i}} \partial_{\ff S_{m}} \ket{\Psi_{j}}) 
- 
\partial_{\ff S_{m}} (\bra{\Psi_{i}} \partial_{S_{k\gamma}} \ket{\Psi_{j}}) \right] 
\dot{S}_{k\gamma}  \nonumber\\
&+ 
i \sum_{k} \sum_{ijl} \sum_{\gamma} 
\left[ \alpha^\ast_l \alpha_j \bra{\Psi_{l}} \partial_{S_{k\gamma}} \ket{\Psi_{i}} - \alpha^\ast_i \alpha_l \bra{\Psi_{j}} \partial_{S_{k\gamma}} \ket{\Psi_{l}} \right] 
\bra{\Psi_{i}} \partial_{\ff S_{m}} \ket{\Psi_{j}}  \dot{S}_{k\gamma} 
\: .
\label{eq:eoms}
\end{align}
The first on the right-hand side is a twofold cross product, $- \dot{\boldsymbol{S}}_m \times ( \nabla_{\boldsymbol{S}_m} \times \boldsymbol{A}_m)$, and with \refeq{mono} and with the normalization $|\ff S_{m}| = 1$, the curl can be written as $\nabla \times \ff A(\ff S) = - \ff S$. 
The second term is an expectation value $\langle \partial_{\ff S_{m}} H \rangle$ of the ``effective field'' $\partial_{\ff S_{m}} H$ in the state of the electron system $|\Psi\rangle$, see \refeq{const}.
With \refeq{con}, the third term reads $\sum_{k} \sum_{ij} \sum_{\gamma} \alpha^\ast_i \alpha_j 
\left[ 
\partial_{S_{k\gamma}} \ff C^{(ij)}_{m}
-
\partial_{\ff S_{m}} C^{(ij)}_{k\gamma}
\right] 
\dot{S}_{k\gamma}
$. 
Its $\beta$-th component involves the ``curl''
\be
\underline{\Omega}^{\rm (A)}_{k\gamma,m\beta} 
= 
\partial_{S_{k\gamma}} \underline{C}_{m\beta}
-
\partial_{S_{m\beta}} \underline{C}_{k\gamma}
\labeq{oma}
\ee
of the spin-Berry connection. 
Here, the underlines indicate that the spin-Berry connection and its curl are matrices in the indices $i,j$ labelling the basis of the low-energy subspace for given spin configuration.
$\underline{\Omega}^{\rm (A)}$ has the form of the spin-Berry curvature in the abelian ($n=1$) theory. 
We refer to this as the ``abelian spin-Berry curvature''.
Again with \refeq{con}, the $\beta$-th component of the fourth term in \refeq{eoms} reads
$
-i \sum_{k} \sum_{ijl} \sum_{\gamma} 
\left[ \alpha^\ast_l \alpha_j C_{k\gamma}^{(li)}  - \alpha^\ast_i \alpha_l C_{k\gamma}^{(jl)} \right] C_{m\beta}^{(ij)}
\dot{S}_{k\gamma}
$. 
This involves the commutator $[\underline C_{k\gamma} , \underline C_{m\beta} ]$ of the spin-Berry connection.

We define the (non-abelian) spin-Berry curvature
\be
\underline{\Omega}_{k\gamma, m\beta}
= 
\partial_{S_{k\gamma}} \underline{C}_{m\beta}
-
\partial_{S_{m\beta}} \underline{C}_{k\gamma}
-
i [ \underline{C}_{k\gamma} , \underline{C}_{m\beta} ]
=
\underline{\Omega}^{\rm (A)}_{k\gamma,m\beta} 
-
i [ \underline{C}_{k\gamma} , \underline{C}_{m\beta} ]
\: ,
\label{eq:curv}
\ee
which differs from the abelian one by the additional commutator.
Furthermore, we define the ``expectation value'' of the spin-Berry curvature in the state given by the wave function $\{\alpha\}$ as:
\be
  \langle {\Omega} \rangle_{k\gamma, m\beta}
  =
  \sum_{ij} \alpha_{i}^{\ast}
  \Omega^{(ij)}_{k\gamma, m\beta}
  \alpha_{j}
\: .  
\labeq{gamma}
\ee
With this, the effective equation of motion (\ref{eq:eoms}) for the classical-spin configuration can be written in the compact form
\be
0 =
\frac{\text{d}}{\text{d} t} \frac{\partial L_{\text{eff}}}{\partial \dot{\boldsymbol{S}}_m} - \frac{\partial L_{\text{eff}}}{\partial \boldsymbol{S}_m} 
= 
\dot{\boldsymbol{S}}_m \times \boldsymbol{S}_m
+
\langle \partial_{\ff S_{m}} \hat{H} \rangle
+ 
\sum_{k} \sum_{\beta\gamma} 
\dot{S}_{k\gamma}
\langle {\Omega} \rangle_{k\gamma, m\beta}
\ff e_{\beta}
\: ,
\ee
or, exploiting the structure of the quantum-classical Hamiltonian \refeq{ham} and the normalization of the wave functions, $\sum_{i} |\alpha_{i}|^{2}=1$,
\be
0
= 
\dot{\boldsymbol{S}}_m \times \boldsymbol{S}_m
+
\langle \partial_{\ff S_{m}} \hat{H}_{\rm int} \rangle
+
\partial_{\ff S_{m}} H_{\rm cl} 
+ 
\sum_{k} \sum_{\beta\gamma} 
\dot{S}_{k\gamma}
\langle {\Omega} \rangle_{k\gamma, m\beta}
\ff e_{\beta}
\: .
\label{eq:eomss}
\ee
This equation is an implicit equation for $\dot{\boldsymbol{S}}_m$. An explicit form is derived in Appendix \ref{sec:exp}. 
Finally, we rewrite \refeq{wf} using the definition of the spin-Berry connection \refeq{con}:
\be
  i \partial_{t} \alpha_{i} 
  = 
  \sum_{j} \bra{\Psi_{i}} ( \hat{H}_{\text{qu}} + \hat{H}_{\text{int}} )\ket{\Psi_{j}} \alpha_{j} 
  - \sum_{m} \sum_{j} \dot{\ff S}_{m} \ff C_{m}^{(ij)} \alpha_{j}
\: .
\labeq{wfs}
\ee
Eqs.\ (\ref{eq:eomss}) and (\ref{eq:wfs}) represent a closed coupled set of non-linear first-order differential equations for the effective many-body wave function $\{\alpha\}$ and for the classical spin configuration $\{ \ff S\}$.

\end{widetext}

\section{Discussion}
\label{sec:dis}

The respective last terms in the equations of motion (\ref{eq:eomss}) and (\ref{eq:wfs}) originate from the strict treatment of the holonomic constraint (\ref{eq:const}).
Although the first time derivative of the local spins is reminiscent of a dissipative Gilbert-like damping, the resulting dynamics is strictly conserving, i.e., the total energy given by the expectation value of the total Hamiltonian (\ref{eq:ham}) with the quantum state of the conduction-electron system is a constant of motion. 
Unlike the standard approach discussed in the introduction, the equations of motion thus describe the dynamics of a closed quantum system (at low energies). 

For the derivation of the equations of motion, we have treated all components of the spins and of the wave function as independent and have thereby disregarded the normalization conditions for the length of the classical spin and for the norm of the wave function
\be
\abs{\boldsymbol{S}_m(t)} =1 \; , \quad \sum_{i}\abs{\alpha_{i}(t)}^{2} = 1 \; ,
\labeq{ncon}
\ee
which must hold at any instant of time $t$. 
One can easily check directly, however, that these are respected.
The normalization condition for the wave function can also be derived by noting that the effective Lagrangian is invariant under global $U(1)$ phase transformations. 
Noether's theorem yields $Q = \sum_{i}\abs{\alpha_{i}(t)}^{2}$ as a conserved charge.
Alternatively, the conditions can be treated as additional constraints via appropriate Lagrange multipliers. 
As is shown in Appendix \ref{sec:norm}, the resulting Euler-Lagrange equations are in fact unchanged. 
	
Adiabatic spin dynamics (ASD) theory \cite{SP17} is recovered for $n=1$, where the conduction-electron dynamics is constrained to the ground-state manifold $\ket{\Psi(t)} = \ket{\Psi_{0}(\{\boldsymbol{S}(t)\})}$ and where the wave function is $\alpha_{0} \equiv 1$ trivially, see \refeq{const}.
In this case, the spin-Berry connection 
$\ff C^{(ij)}_{m} = i \bra{\Psi_{i}} \partial_{\ff S_{m}} \ket{\Psi_{j}}$ with $i,j=0,...,n-1$, reduces to a vector with scalar entries only, 
$\ff C_{m} = i \bra{\Psi_{0}} \partial_{\ff S_{m}} \ket{\Psi_{0}}$. 
Hence, the commutator in \refeq{curv} vanishes, and the spin-Berry curvature $\underline{\Omega}_{k\gamma, m\beta}$ reduces to the corresponding expression $\underline{\Omega}^{\rm (A)}_{k\gamma,m\beta}$, \refeq{oma}, of (abelian) ASD theory. 

In the opposite extreme case, i.e., when $n$ is chosen as the dimension of the full many-electron Fock space $\cal H$, \refeq{const} is actually no longer a constraint but rather represents the expansion of the electron state $|\Psi(t)\rangle$ with respect to a complete orthonormal system of time-dependent basis states $\{ | \Psi_{i}(t) \rangle \}$ with $|\Psi_{i}(t) \rangle=|\Psi_{i}(\{ \ff S(t) \} \rangle$. 
In this case, it is straightforward to see that \refeq{wfs} is just Schr\"odingers's equation $i\partial_{t}|\Psi(t)\rangle=\hat{H} |\Psi(t)\rangle$, i.e., \refeq{eom1}, but formulated for the coefficients $\alpha_{i}(t)$ of $|\Psi(t)\rangle$ in that basis. 
The spin-Berry connection merely takes care of the fact that the basis changes smoothly with the parameters $\{ \ff S \}$. 
\refeq{eomss} trivializes as well in this case: 
We can rewrite the (non-abelian) spin-Berry curvature in the form (see Appendix \ref{sec:proj}):
\be
\Omega^{(ij)}_{k\gamma,m \beta} 
= 
i
\left[
\bra{\partial_{S_{k\gamma}}\Psi_{i}}\mathcal{Q}_{n}\ket{\partial_{S_{m\beta}}\Psi_{j}} 
- 
(
k\gamma  \leftrightarrow m\beta
)
\right]
\labeq{sbcq}
\: ,
\ee
where $\mathcal{Q}_{n} := \mathbb{1}-\sum_{i=0}^{n-1}\ket{\Psi_{i}}\bra{\Psi_{i}}$ projects onto the orthogonal complement of the low-energy space ${\cal E}_{n}(\{ \ff S \})$.
If $n = \dim \cal H$, the complement is zero, and the spin-Berry curvature vanishes identically, so that
\be
0
= 
\dot{\boldsymbol{S}}_m \times \boldsymbol{S}_m
+
\langle \partial_{\ff S_{m}} \hat{H}_{\rm int} \rangle
+
\partial_{\ff S_{m}} H_{\rm cl} 
\: .
\label{eq:dimn}
\ee
Taking the cross product with $\ff S_{m}$ from the right on both sides of \refeq{eomss} and exploiting the normalization condition for the spin length, we get:
\be
\dot{\boldsymbol{S}}_m 
=
\frac
{\partial \hat{H} (\{\boldsymbol{S}\})}{\partial \ff S_{m}} \times \ff S_{m} \: .
\labeq{eomsfull}
\ee
This is just the explicit form of \refeq{eom2}. 

Some general properties of the spin-Berry curvature can be derived from \refeq{sbcq}. 
One immediately notes the antisymmetry
\be
\Omega^{(ij)}_{k\gamma,m \beta} 
=
-
\Omega^{(ij)}_{m\beta, k\gamma} 
\labeq{prop1}
\ee
for fixed $i,j$.
Furthermore, complex conjugation yields
\be
\Omega^{(ij)^{*}} _{k\gamma,m \beta}
=
-
\Omega^{(ji)}_{m\beta, k\gamma} 
\: .
\labeq{prop2}
\ee
With these properties, one can immediately conclude that 
\be
  \langle {\Omega} \rangle_{k\gamma, m\beta}
  =
  \sum_{ij} \alpha_{i}^{\ast}
  \Omega^{(ij)}_{k\gamma, m\beta}
  \alpha_{j}
  =
  \langle {\Omega} \rangle_{k\gamma, m\beta}^{\ast}
\: ,   
\labeq{omreal}
\ee
i.e., the expectation value, which enters the effective equation of motion \refeq{eomss}, is real. 

Quite generally, the (abelian) Berry connection and Berry curvature arise in the adiabatic problem, where a quantum Hamiltonian $\hat{H}=\hat{H}(\ff \lambda)$ depends on a family of slowly varying parameters $\ff \lambda$ and has a non-degenerate ground state for all $\ff \lambda$. 
This gives rise to the famous Berry phase \cite{Ber84}, which the ground state picks up during a closed loop in parameter space and which can be computed, e.g., as an integral of the Berry curvature over the surface bounded by the loop.
Mathematically, the phase is a holonomy, i.e., it results from a twist of the line bundle $\{ ( \ff \lambda, | \Psi_{0} \rangle ) \, | \, \hat{H}(\ff \lambda) | \Psi_{0} \rangle = E_{0}(\{ \ff \lambda \}) | \Psi_{0} \rangle \}$ \cite{Sim83}.
The Berry phase is gauge invariant and thus observable and depends on the geometry of the closed loop only.
Similarly, non-abelian gauge fields arise in the adiabatic time evolution of an $n>1$-fold degenerate ground state of a quantum system \cite{WZ84} and produce a non-trivial phase after completing a loop in parameter space. 

Here, we consider a quantum system coupled to {\em dynamical} classical degrees of freedom (classical spins). 
In case of a clear time-scale separation between the slow classical and the fast quantum dynamics, the classical spins induce a spin-Berry curvature in the quantum conduction-electron system. 
Generically, it is highly unlikely, however, that the classical state evolves along a closed path. 
The essential observation, however, is that there is an additional {\em feedback} of the Berry curvature on the classical spin dynamics, seen in the last term in \refeq{eomss} for $\Omega=\Omega^{\rm (A)}$.
Already in the abelian case $n=1$, this leads to an anomalous geometrical spin torque \cite{SP17}.
This geometric feedback on slow classical dynamics has been pointed out \cite{WZ88,NK98,BMK+03,NWK+99,SP17,EMP20,BN20,MP21} but has not yet been studied for spin dynamics in the non-abelian case $1< n = \dim {\cal E}_{n}\{ \ff S\} \ll \dim \ca H$.

\section{Time reversal}
\label{sec:tr}

Time-reversal symmetry plays an important role for the presence of a finite spin-Berry curvature in the adiabatic case ($n=1$) \cite{SP17}.
For $n>1$, however, this is entirely different:

We assume that the electron system is time-reversal symmetric, i.e., that the Hamiltonian $\hat{H}_{\text{qu}}$ commutes with the anti-unitary operator for time reversal $\Theta$.
The interaction term, \refeq{hint}, on the other hand, is odd under time reversal, $\Theta \hat{H}_{\rm int} \Theta^{\dagger} = - \hat{H}_{\rm int}$, since $\Theta \ff s_{\ff r_{m}} \Theta^{\dagger} = - \ff s_{\ff r_{m}}$.
The local spins $\ff S_{m}$ are classical degrees of freedom, which act as local magnetic fields and explicitly break time-reversal symmetry of the quantum system.

This effect, however, can be disregarded in the weak-$J$ regime, where the spin-Berry curvature, in the spirit of linear-response theory, is a physical property of the electron system $\hat{H}_{\text{qu}}$ only. 
Namely, expanding $E_{i} = E_{i0} + \ca O (J)$ and $| \Psi_{l} \rangle = | \Psi_{l}^{0} \rangle + {\cal O}(J)$ and using the identity
\be
  \langle \Psi_{l} | \partial_{\ff S_{m}} \Psi_{j} \rangle 
  = 
  \frac{\langle \Psi_{l} | \partial_{\ff S_{m}} \hat{H} (\{\boldsymbol{S}\}) | \Psi_{j} \rangle }{E_{j} - E_{l}} \: , 
\ee
which holds for $E_{j} \ne E_{l}$, \refeq{sbcq} can be rewritten as
\ba
\Omega^{(ij)}_{k\gamma,m \beta} 
& = &
i \sum_{l \ge n}
\Bigg[
\frac{\langle \Psi^{0}_{i} | \partial_{S_{k\gamma}} \hat{H} | \Psi_{l}^{0} \rangle }{E_{i0} - E_{l0}}
\frac{\langle \Psi^{0}_{l} | \partial_{S_{m    \beta}} \hat{H} | \Psi_{j}^{0} \rangle }{E_{j0} - E_{l0}}
\nonumber \\
& - &
(
k\gamma  \leftrightarrow m\beta
)
\Bigg] + {\cal O}(J^{3})
\: , 
\labeq{sbcl}
\ea
since $\partial_{S_{k\gamma}} \hat{H}_{\rm int} = J s_{\rm i_{k}\gamma} = {\cal O}(J)$, so that the spin-Berry curvature is of order $J^{2}$ for weak $J$ and expressed in terms of the eigenstates and eigenenergies of $\hat{H}_{\text{qu}}$ only. 
Note that $0\le i,j \le n-1$ in \refeq{sbcl}.

For a system with an even number of spin-$1/2$ electrons, the time-reversal operator squares to unity, $\Theta^{2}=+ 1$. 
In this case, we can choose an orthonormal basis of time-reversal-symmetric energy eigenstates  $| \Psi^{0}_{i} \rangle = \Theta  | \Psi^{0}_{i} \rangle$. 
This implies that the matrix elements,
\ba
\langle \Psi^{0}_{i} | \partial_{S_{k\gamma}} \hat{H} | \Psi^{0}_{l} \rangle
&=&
- \langle \Psi^{0}_{i} | \Theta^{\dagger} \partial_{S_{k\gamma}} \hat{H} \Theta | \Psi^{0}_{l} \rangle
\nonumber \\
&=&
- (\langle \Theta \Psi^{0}_{i} | \partial_{S_{k\gamma}} \hat{H} | \Theta \Psi^{0}_{l} \rangle)^{\ast}
\nonumber \\
&=&
- (\langle \Psi^{0}_{i} | \partial_{S_{k\gamma}} \hat{H} | \Psi^{0}_{l} \rangle)^{\ast} 
\; ,
\labeq{me}
\ea
are purely imaginary. 
Note that only the (odd) interaction term $\hat{H}_{\rm int}(\{\ff S\})$ contributes. 
Using this in \refeq{sbcl} shows that $\Omega^{(ij)}_{k\gamma,m \beta}$ is purely imaginary. 
With \refeq{prop2} we can conclude that
\be
\Omega^{(ij)} _{k\gamma,m \beta}
=
\Omega^{(ji)}_{m\beta, k\gamma} 
\: .
\labeq{prop3}
\ee
In particular, \refeq{prop1} and \refeq{prop3} imply that the $i=j$ elements of the spin-Berry curvature must vanish in the weak-$J$ limit for $\Theta^{2}=+1$.
This is important for the abelian case $n=1$. 
For $i=j=0$ we have $\Omega^{(00)} _{k\gamma,m \beta}=0$ and, hence, there is no geometrical spin torque in the weak-$J$ limit for a time-reversal-symmetric system with $\Theta^{2}=+1$.
In the general non-abelian case, on the other hand, we find with \refeq{omreal} that
\be
  \langle {\Omega} \rangle_{k\gamma, m\beta}
  =
  - \sum_{ij} \mbox{Im}(\alpha_{i}^{\ast}  \alpha_{j})
  \mbox{Im}\Omega^{(ij)}_{k\gamma, m\beta}
\: , 
\labeq{omnonzero}
\ee
since $\Omega^{(ij)} _{k\gamma,m \beta}$ is imaginary. 
Generically, the coefficients $\alpha_{i} = \alpha_{i}(t)$ in the expansion \refeq{const} will be complex and oscillatory functions of time. 
The expression above thus shows that even in the weak-$J$ limit and for a time-reversal symmetric system, the geometrical spin torque in the equation of motion (\ref{eq:eomss}) is generally finite.

Let us briefly discuss the case of an odd electron number with $\Theta^{2}=-1$. 
Here, the basis states can be grouped in orthogonal and energy-degenerate Kramers pairs $\{ | \Psi^{0}_{i} \rangle , | \overline{\Psi}_{i}^{0} \rangle\}$ with $| \overline{\Psi}^{0}_{i} \rangle \equiv \Theta | \Psi_{i}^{0} \rangle$ for $i=0,...,(n/2)-1$. 
An even number of states must be included in formulating the constraint (\ref{eq:const}).
For the matrix elements, we have
\ba
\langle \Psi^{0}_{i} | \partial_{S_{k\gamma}} \hat{H} | \Psi^{0}_{l} \rangle
&=&
- \langle \Psi^{0}_{i} | \Theta^{\dagger} \partial_{S_{k\gamma}} \hat{H} \Theta | \Psi^{0}_{l} \rangle
\nonumber \\
&=&
- (\langle \Theta \Psi^{0}_{i} | \partial_{S_{k\gamma}} \hat{H} | \Theta \Psi^{0}_{l} \rangle)^{\ast}
\nonumber \\
&=&
- (\langle \overline{\Psi}^{0}_{i} | \partial_{S_{k\gamma}} \hat{H} | \overline{\Psi}^{0}_{l} \rangle)^{\ast}
\; .
\labeq{me2}
\ea
This can be used in \refeq{sbcl} since in the $l$-sum with each term also the Kramers partner is included. 
We find
\be
\Omega^{(ij)}_{k\gamma,m \beta}
=
\Omega^{(\overline{j} \, \overline{i})}_{m\beta, k\gamma} 
\: ,
\labeq{prop3a}
\ee
where the index $\overline{i}$ refers to the Kramers partner of $| \Psi^{0}_{i} \rangle$, and, 
furthermore, $(\Omega^{(ij)}_{k\gamma,m \beta})^{\ast} = - \Omega^{(\overline{i} \, \overline{j})}_{k\gamma,m \beta}$.
As for the case $\Theta^{2}=+1$, time-reversal symmetry does not lead to a vanishing spin-Berry curvature or a vanishing expectation value 
$\langle {\Omega} \rangle_{k\gamma, m\beta}$.
Note that for $\Theta^{2}=-1$ the adiabatic theory is not applicable anyway (for the weak-coupling limit), since the ground state is at least twofold Kramers degenerate.

\section{Gauge transformations}
\label{sec:gauge}

The effective Lagrangian \refeq{leff} can be written in a compact form as 
\ba
L_{\text{eff}} 
& = &
\sum_{m}
\boldsymbol{A}_{m}(\ff S_{m}) \dot{\boldsymbol{S}}_{m} 
+ 
i \boldsymbol{\alpha}^{\dagger}\partial_{t}\boldsymbol{\alpha} 
\nonumber \\
&+& 
\sum_{m}\boldsymbol{\alpha}^{\dagger} [ \underline{\ff C}_{m}(\{\ff S\}) \dot{\boldsymbol{S}}_{m} ] \boldsymbol{\alpha} - \boldsymbol{\alpha}^{\dagger}\underline{H}(\{ \ff S\})\boldsymbol{\alpha} 
\; ,
\labeq{leffc}
\ea
where $\ff \alpha = (\alpha_{0}, ..., \alpha_{n-1})^{T}$ and where $\underline{H}$ is the Hamilton matrix with elements $H_{ij} = \bra{\Psi_{i}}\hat{H}\ket{\Psi_{j}}$ and the local basis states $|\Psi_{j} \rangle = | \Psi_{j} (\{ \ff S\}) \rangle$.
We consider a gauge transformation
\ba
  | \Psi_{j}(\{ \ff S\}) \rangle
  & \mapsto &
  | \Psi'_{j}(\{ \ff S\}) \rangle
  =
  \sum_{i}
  U_{ij}^{\dagger} 
  | \Psi_{i}(\{ \ff S\}) \rangle
\nonumber \\
  \ff \alpha 
  & \mapsto &
  \ff \alpha'
  = 
  \underline{U} 
  \ff \alpha  
  \: , 
\labeq{rot}  
\ea
where $\underline U$ (with elements $U_{ij}$) is the defining matrix representation of SU(n) on the local low-energy subspace $\ca E_{n} (\{ \ff S\})$ for given spin configuration $\{ \ff S\}$. 
This transformation must leave observables invariant, since \refeq{rot} just means a rotation of the basis in $\ca E_{n} (\{ \ff S\})$, which leaves the quantum state $| \Psi \rangle = \sum_{j=0}^{n-1} \alpha_{j} | \Psi_{j} (\{ \ff S\}) \rangle$, and thus the constraint \refeq{const} invariant when rotating the expansion coefficients (the wave function) accordingly.
We distinguish between global SU(n) and local SU(n) transformations. 
For the latter, the transformation matrix $\underline U = \underline U(\{ \ff S \})$ is an arbitrary but smooth function of the spin configuration $\{ \ff S\}$.
The effective Lagrangian is invariant under both, global and local gauge transformations. 

Note that the Hamilton matrix transforms in a covariant way,
\be
\underline{H} 
\mapsto
\underline{H}^\prime = \underline{U}\, \underline{H}\, \underline{U}^\dagger   
\labeq{hgauge}
\: , 
\ee
while the Berry connection transforms covariantly under a global gauge transformation only. 
For a local gauge transformation we rather have:
\be
\underline{\ff C}_m
\mapsto
\underline{\ff C}_{m}^\prime = \underline{U}\, \underline{\ff C}_m\, \underline{U}^\dagger + i \underline{U} \partial_{\ff S_{m}} \underline{U}^\dagger 
\: . 
\labeq{cgauge}
\ee
The non-abelian Berry curvature, opposed to its abelian part (\ref{eq:oma}), transforms covariantly:
\be
\underline{\Omega}_{k\gamma, m\beta}
\mapsto
\underline{\Omega}'_{k\gamma, m\beta}
=
\underline{U}\,
\underline{\Omega}_{k\gamma, m\beta}\,
\underline{U}^{\dagger}\,
\: ,
\labeq{bgauge}
\ee
so that its expectation value in the state given by the wave function $\alpha_{i}$ is invariant:
$\langle {\Omega'} \rangle'_{k\gamma, m\beta} = \langle {\Omega} \rangle_{k\gamma, m\beta}$. 
Hence, \refeq{eomss} is invariant under local gauge transformations. 
The Schr\"odinger-type equation \refeq{wfs}, on the other hand, is form-invariant under local transformations, i.e.,
\be
  i \partial_{t} \alpha'_{i} 
  = 
  \sum_{j} \bra{\Psi'_{i}} ( \hat{H}_{\text{qu}} + \hat{H}_{\text{int}} )\ket{\Psi'_{j}} \alpha'_{j} 
  - \sum_{mj} \dot{\ff S}_{m} {\ff C_{m}^{(ij)}}'\alpha'_{j}
\: ,
\labeq{wfstrans}
\ee
and the spin-Berry connection term on the right-hand side is necessary to compensate the extra term appearing on the left-hand side in case of an $\{ \ff S\}$-dependent transformation. 

Concluding, the effective Lagrangian emerging in the low-energy sector of hybrid spin-electron dynamics represents a non-abelian SU(n) gauge theory.
This is reminiscent of standard quantum field theories \cite{PS96}, where the Lagrangian is invariant under simultaneous transformations of coupled matter and gauge fields, and where these gauge transformations involve a gauge group, like SU(n), and are local in space-time.
There are a couple of differences though:
Within non-abelian spin-dynamics theory, space-time is not only replaced by a compact parameter manifold, namely the Cartesian product of classical Bloch spheres representing the space of the spin configurations, but furthermore the spin configurations have their own dynamics. 
The theory is thus much more related to gauge theories that have been devised for molecular physics \cite{BMK+03}, where the state space of the nuclei, when treated classically, define a dynamical parameter manifold, and where the role of the gauge field is played by the non-abelian Berry connection. 

Finally, it is worth mentioning that there is a second, less important class of gauge freedom. 
This concerns the vector potential $\ff A(\ff S_{m})$, see the first term of $L$ in \refeq{lag}, i.e., already in the {\em full} Lagrangian.
Any transformation of the unit vector $\ff e \mapsto \ff e'$ leads to a transformed potential $\ff A(\ff S_{m}) \mapsto \ff A'(\ff S_{m})$ but leaves its curl invariant.
This even includes ``local'', $m$-dependent transformations $\ff A(\ff S_{m}) \mapsto \ff A'_{m}(\ff S_{m})$ resulting from $\ff e\mapsto \ff e'_{m}$.  
However, since only the curl $\nabla_{\ff S} \times \ff A(\ff S_{m})$ enters the equations of motion resulting from the full or from the effective Lagrangian, see \refeq{eoms} for instance, these are invariant.

\section{Minimal model}
\label{sec:mod}

\begin{figure}[t]
\includegraphics[width=0.6\columnwidth]{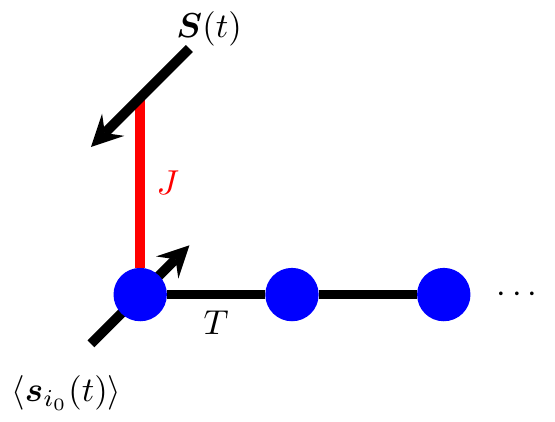}
\caption{
Sketch of the minimal model studied numerically. 
A classical spin $\ff S$ of length $|\ff S|=1$ is antiferromagnetically exchange coupled with coupling strength $J>0$ to the local spin moment $\ff s_{i_{0}}$ at the first site $i_{0}=1$ of a system of conduction electrons on a one-dimensional chain with open boundaries.
$T$ is the nearest-neighbor hopping.
Real-time dynamics is initiated by a sudden change of the direction of a local magnetic field $\ff B$ coupled to $\ff S$.
}
\label{fig:sys}
\end{figure}

For a further discussion of non-abelian spin-dynamics theory, we will present numerical results for a minimal model, which includes a few degrees of freedom only but is sufficient to illustrate several key aspects.
Our intention is to show by example that and how our theoretical approach can be evaluated in practice, how the numerical results compare with the full solution of the equations of motion, and what improvements the theory offers over the purely adiabatic (abelian) version. 
This may also be seen as a preparation for future applications to more realistic but also more complicated physical systems, where various secondary issues become important.

The Hamiltonian our our toy model is given by
\be 
  \hat{H} = -T \sum_{\langle i,j \rangle,\sigma} c_{i\sigma}^{\dagger} c_{j\sigma}
  + J \boldsymbol{s}_{i_{0}}\boldsymbol{S}-\boldsymbol{B}\boldsymbol{S}
\: .
\labeq{min}
\ee
It describes a single classical spin ($M=1$) locally exchange coupled (coupling constant $J>0$) to a non-interacting tight-binding model in an open chain geometry with a small number of sites $L$ hosting $N=L$ electrons, i.e., a half-filled conduction-electron system. 
The spin is coupled to the first site of the chain $i_{0}=1$. 
This is the $s$-$d$ model \cite{vz} discussed in the introduction and the same model as in Ref.\ \cite{SP17}.
Energy and time units are fixed by setting the nearest-neighbor hopping amplitude to $T=1$.
In addition, the Hamiltonian includes a local magnetic field of strength $B$ coupling to the classical spin $\ff S$. 
The model is visualized in Fig.\ \ref{fig:sys}. 

The field term is employed to initiate the real-time dynamics:
At time $t=0$ the system is prepared in the ground state of $\hat{H}$ with the field in $x$ direction, i.e., the spin $\ff S = S \ff e_{x}$ is aligned to $\ff B = B \ff e_{x}$, and the conduction-electron state is the ground state $| \Psi(t=0) \rangle = | \Psi_{0} (\ff S) \rangle$.
Time propagation for $t>0$ is driven by the same Hamiltonian but with the field pointing in $z$ direction. 
Dynamics is thus initiated by a sudden change of the field direction from $x$- to $z$ direction.

For $t>0$ one expects that the spin starts precessing around the $z$ axis. 
In the adiabatic approximation with $n=1$, the electron system will follow the respective spin direction instantaneously, and its state at time $t$ would be the instantaneous ground state $| \Psi_{0} (\ff S(t)) \rangle$.
The time scale on which the precession takes place is given by the inverse of the Larmor frequency $\omega_{\rm L} = B$. 
Depending on the field strength, this time scale $\tau_{L} = 1/\omega_{\rm L} = 1/B$ can be much shorter than the inverse of the finite-size gap $\Delta = \ca O(T/L)$. 
With $T=1$ we thus expect that the adiabatic approximation breaks down for $B \gg T/L$ and that excited states $| \Psi_{j} (\ff S) \rangle$ with $0<j<n-1$ will be populated. 
The number of states $n$ included in the $\ff S$-dependent basis controls the accuracy of the non-abelian spin-dynamics approach. 

For the single-classical-spin model the effective equations of motion \refeq{eomss} and \refeq{wfs} are somewhat simplified. 
For $M=1$ we can skip the $m$-index and take the cross product with $\ff S$ on both sides of \refeq{eomss}. 
Furthermore, we have 
$\langle \partial_{\ff S} \hat{H}_{\rm int} \rangle = J \langle \ff s_{i_{0}} \rangle - \ff B$
and 
$\partial_{\ff S} H_{\rm cl} = 0$. 
Therewith we get
\be		
\dot{\boldsymbol{S}} = \frac{J\langle\boldsymbol{s}_{i_{0}}\rangle \times \boldsymbol{S}-\boldsymbol{B}\times\boldsymbol{S}}{1 - \boldsymbol{S} \langle \boldsymbol{\Omega} \rangle}
\: , 
\labeq{ren}
\ee
where
$\langle \boldsymbol{\Omega} \rangle = \sum_{ij} \alpha_{i}^{\ast} \ff \Omega^{(ij)} \alpha_{j}$ is the expectation value of the pseudovector $\ff \Omega^{(ij)}$ with components $\Omega^{(ij)}_{\alpha} = \frac12 \sum_{\beta\gamma} \varepsilon_{\alpha\beta\gamma}\Omega^{(ij)}_{\beta\gamma} $ that can be constructed for $M=1$ due to the antisymmetry of the Berry curvature tensor under $\beta \leftrightarrow \gamma$ for each pair $(ij)$, see \refeq{prop1}.
Furthermore, 
$\langle\boldsymbol{s}_{i_{0}}\rangle = \langle\boldsymbol{s}_{i_{0}}\rangle_{t} = \sum_{ij} \alpha_{i}^{\ast}(t) \langle \Psi_{i}(\ff S(t)) | \ff s_{i_{0}} | \Psi_{j}(\ff S(t)) \rangle \alpha_{j}(t)$.

Remarkably, there is a renormalization of the precession frequency resulting from the geometrical spin torque, which has already been studied for the adiabatic case \cite{SP17,BN20,EMP20,MP21}.
This manifests itself as an additional factor $1 / ( 1 - \boldsymbol{S} \langle \boldsymbol{\Omega} \rangle )$ in \refeq{ren}.
In the adiabatic case $n=1$, the expectation value $\langle \boldsymbol{\Omega} \rangle$ is strictly parallel or antiparallel to classical-spin orientation due to symmetry reasons \cite{SP17}. 
For $\boldsymbol{S} \uparrow \uparrow \langle \boldsymbol{\Omega} \rangle$ this results in a faster precessional dynamics, and its orientation is even reversed if $\boldsymbol{S} \langle \boldsymbol{\Omega} \rangle > 1$, while for $\boldsymbol{S} \uparrow \downarrow \langle \boldsymbol{\Omega} \rangle$ the precession is slowed down.
Exactly at $\boldsymbol{S} \langle \boldsymbol{\Omega} \rangle = 1$ the right-hand side of \refeq{ren} becomes singular.
This is linked to a divergence of the precession frequency which, however, becomes relevant in an extreme case only: 
For the adiabatic case and $L=1$, it was found in Ref.\ \cite{SP17} that singular dynamics can in principle be approached, if the length of the classical spin $|\ff S| \to \frac12$. 
At the same time, however, to stay in the adiabatic regime of the model, it was necessary to consider an ever-increasing coupling strength, i.e., $J\to \infty$. 

Here, we see that the same type of singularity is in principle also present in the non-adiabatic case (for $M=1$).
Generally, however, we find $0 < \boldsymbol{S} \langle \boldsymbol{\Omega} \rangle<1$ (for antiferromagnetic exchange coupling $J>0$):
A possible singularity is regularized for $n>1$ due to contributions from excited states and partly also to due the fact that $\langle \ff \Omega \rangle$ and $\ff S$ are no longer necessarily collinear. 

The following NA-SD studies of the minimal model are based on a numerical solution of the coupled effective equations of motion \refeq{ren} for the classical spin $\ff S$ and \refeq{wfs} for the wave function $\{\alpha\}$. 
For the computation of the expectation value of the spin-Berry curvature $\langle \ff \Omega \rangle$ we profit from simplifications, which hold in case of a non-interacting conduction-electron system. 
These are detailed in the Appendix \ref{sec:comp}.

We also compare the results of the NA-SD theory with the {\em full} solution of the fundamental equations of motion (\ref{eq:eom1}) and (\ref{eq:eom2}), which is obtained independently. 
More explicitly, \refeq{eom1} for the minimal model reads:
\be
\dot{\boldsymbol{S}} = J\langle\boldsymbol{s}_{i_{0}}\rangle_{t}\times\boldsymbol{S}-\boldsymbol{B}\times\boldsymbol{S}
\: .
\labeq{e1}
\ee
Furthermore, in case of a non-interacting electron system, \refeq{eom2} can be replaced by the equation of motion 
\be
i \frac{d}{dt} \underline{\rho} = \comm{\underline{T}^{(\text{eff})}}{\underline{\rho}}
\labeq{e2}
\ee
for the one-particle reduced density matrix $\underline{\rho}$ with elements $\rho_{ii'\sigma\sigma^{\prime}}(t)\ = \langle c^{\dagger}_{i'\sigma^{\prime}}c_{i\sigma}\rangle$, and where the elements of the effective hopping matrix $\underline{T}^{(\text{eff})}$ are given by: 
\be
T^{(\text{eff})}_{ii'\sigma\sigma^{\prime}} 
= 
T\delta_{\langle ii' \rangle} \delta_{\sigma\sigma^{\prime}}
+
\frac{J}{2} \ff \sigma_{\sigma\sigma^{\prime}} \ff S \, \delta_{ii_{0}} \delta_{i'i_{0}}
\; .
\labeq{teff}
\ee

\section{Numerical results}
\label{sec:res}

\subsection{Full theory}

\begin{figure}[b]
\centering
\includegraphics[width=0.95\columnwidth]{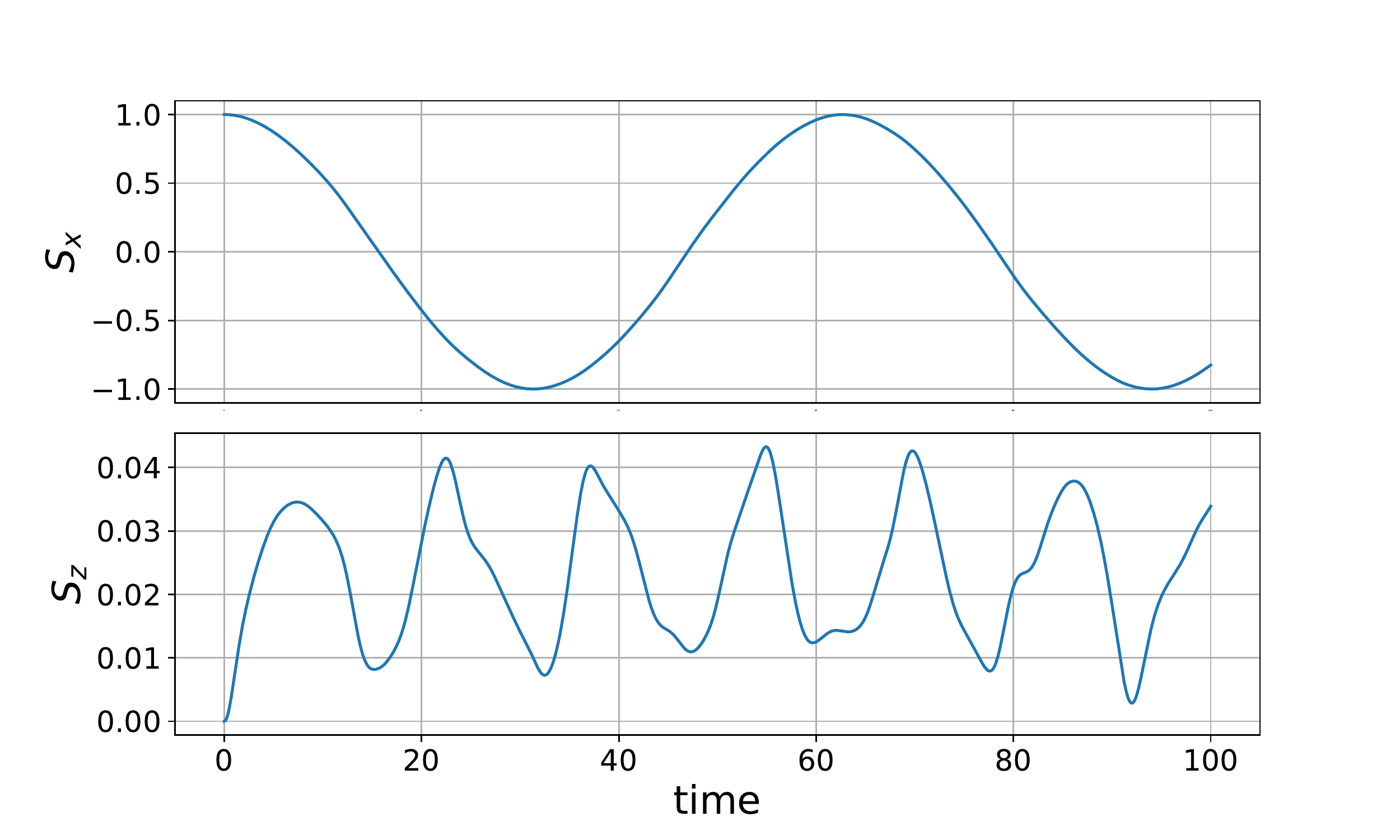}
\caption{
Time evolution of the $x$ and the $z$ component of the classical spin as obtained from the full theory for a system with $L=10$ sites at half-filling $N=L$.
Parameters: $J=1$, $B=0.1$. 
The energy and time units are set by fixing the nearest-neighbor hopping at $T=1$.
}
\label{fig:prec}
\end{figure}

The precession around the $z$ axis defined by the local magnetic field is expected to be the dominant effect in the classical spin dynamics. 
In fact, this is the main phenomenon found by solving the full set of equations of motion (\ref{eq:e1}) and (\ref{eq:e2}). 
Fig.\ \ref{fig:prec} displays numerical results obtained with the full theory for a system with $L=10$ sites at half-filling $N=L$, and for generic parameter values $J=1$ and $B=0.1$.
The $x$ component of the classical spin undergoes a quite regular oscillation with a period close to $2\pi / \omega_{\rm L} = 2\pi /B \approx 62.8$. 
The $y$ component exhibits the same but phase shifted dynamics. 
We note that, for the selected parameter set, the geometrical spin torque is too small to produce a sizeable renormalization of the precession frequency.

Damping of the spin dynamics and eventual alignment of the classical spin with the field $\ff B=B \ff e_{z}$ is typically a weaker effect, which takes place on a much longer time scale, see e.g.\ the discussion in Refs.\ \cite{llg,BNF12,SP15}.
For a closed, finite and with $L=10$ small system, as considered here, relaxation will be imperfect anyway, and even in the long-time limit, the system cannot fully approach its ground state locally, in the vicinity of $i_{0}$.
Uncovering this type of relaxation dynamics requires much larger systems, as discussed in Refs.\ \cite{EP20,EP21}, for example.

Fig.\ \ref{fig:prec} also displays the $z$ component of the spin. 
In case of a perfect precessional motion, one would expect a constant $S_{z}$. 
As is seen in the figure, however, an almost oscillatory motion of $S_{z}$ with some additional irregularities is found instead. 
This nutation of the spin is reminiscent of gyroscope theory \cite{But06,WC12}, but is not understood easily.
An explanation in terms of linear-response theory (see Eq.\ (\ref{eq:integro})), i.e., Redfield theory for open quantum systems, involves the second-order term in the Taylor expansion of the memory kernel \cite{BNF12,SRP16b}.
For the parameters considered here, the nutation effect is at least an order of magnitude smaller as compared to the precessional dynamics (see Fig.\ \ref{fig:prec}). 
There are cases, however, where precessional and nutational oscillations can be of the same order of magnitude.
The additional ``irregularities'' on top of the nutation are even more subtle. 
At this level of resolution at the latest, the complexity of the dynamics caused by the nonlinearity of the quantum-classical equations of motion appears to prohibit a simple explanation.

\subsection{Anomalous precession}

\begin{figure}[b]
	\centering
	\includegraphics[width=0.95\columnwidth]{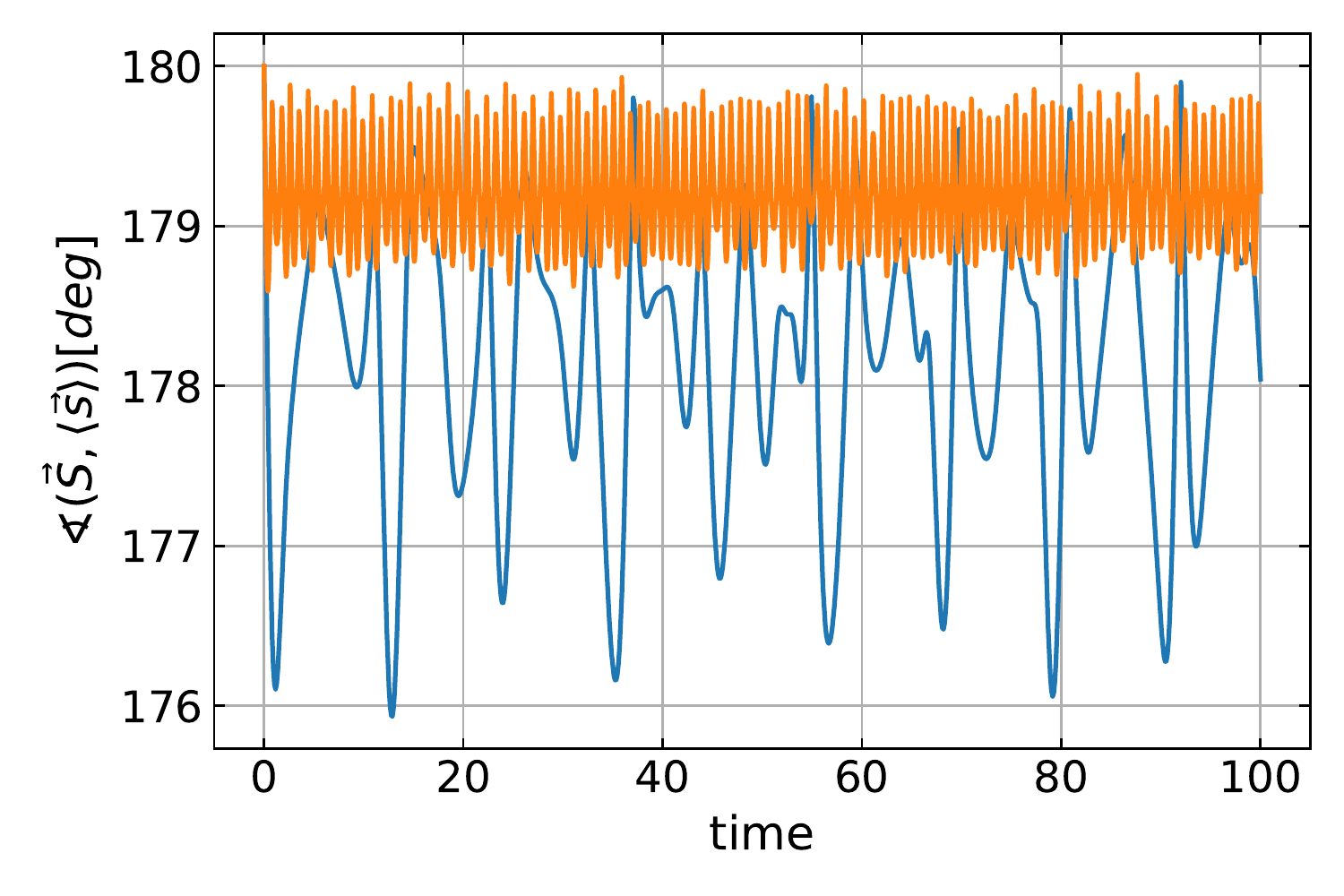}
\caption{
Time evolution of the angle enclosed by the classical spin $\boldsymbol{S}$ and the expectation value of the local spin of the electron system at the impurity site $\langle \boldsymbol{s}_{i_0} \rangle$. 
Results as obtained by the full theory for $J=1$ (blue) and $J=15$ (orange). 
Other parameters as in Fig.\ \ref{fig:prec}.
}
\label{fig:angle}
\end{figure}

In the case of strong exchange coupling $J \gg T$, the classical spin $\ff S$ and the local magnetic moment $\langle \ff s_{i_{0}} \rangle$ at $i_{0}$ are tightly bound together. 
In this regime one would thus expect that $\langle \ff s_{i_{0}} \rangle$ follows the classical-spin direction almost instantaneously such that $\langle \ff s_{i_{0}}\rangle$ is almost perfectly aligned antiferromagnetically to $\ff S$.
The time evolution of the angle enclosed by $\ff S$ and $\langle \ff s_{i_{0}}\rangle$ is shown in Fig.\ \ref{fig:angle}. 
For $J=1$ the mean deviation of the angle from $180^{\circ}$ is in fact about $2^{\circ}$ only, and it shrinks with increasing $J$, see the result for $J=15$. 
On the other hand, the absolute value of the local moment $\langle \ff s_{i_{0}} \rangle$ of the conduction-electron systems that is induced by $\ff S$, increases from $|\langle \ff s_{i_{0}} \rangle| \approx 0.18$ at $J=1$ to $|\langle \ff s_{i_{0}} \rangle| \approx 0.49$ at $J=15$.
The net effect, however, is that the spin torque on $\ff S$ originating from the exchange term, $J \langle \ff s_{i_{0}}\rangle \times \ff S$ is weak compared to the torque due to the field $- \ff B \times \ff S$.
Following naive adiabatic theory one would therefore expect a precessional motion of $\ff S$ in the $x$-$y$ plane with a frequency $\omega_{\rm p}$ close to the Larmor frequency $\omega_{\rm L} = B$.
However, this naive picture in principle disregards the effect due to the geometrical spin torque, which can be sizeable. 
It is thus instructive to compare the naive expectation as well as adiabatic spin dynamics theory (ASD) with the full solution of the fundamental equations of motion.

\begin{figure}[t]
\centering
\includegraphics[width=0.95\columnwidth]{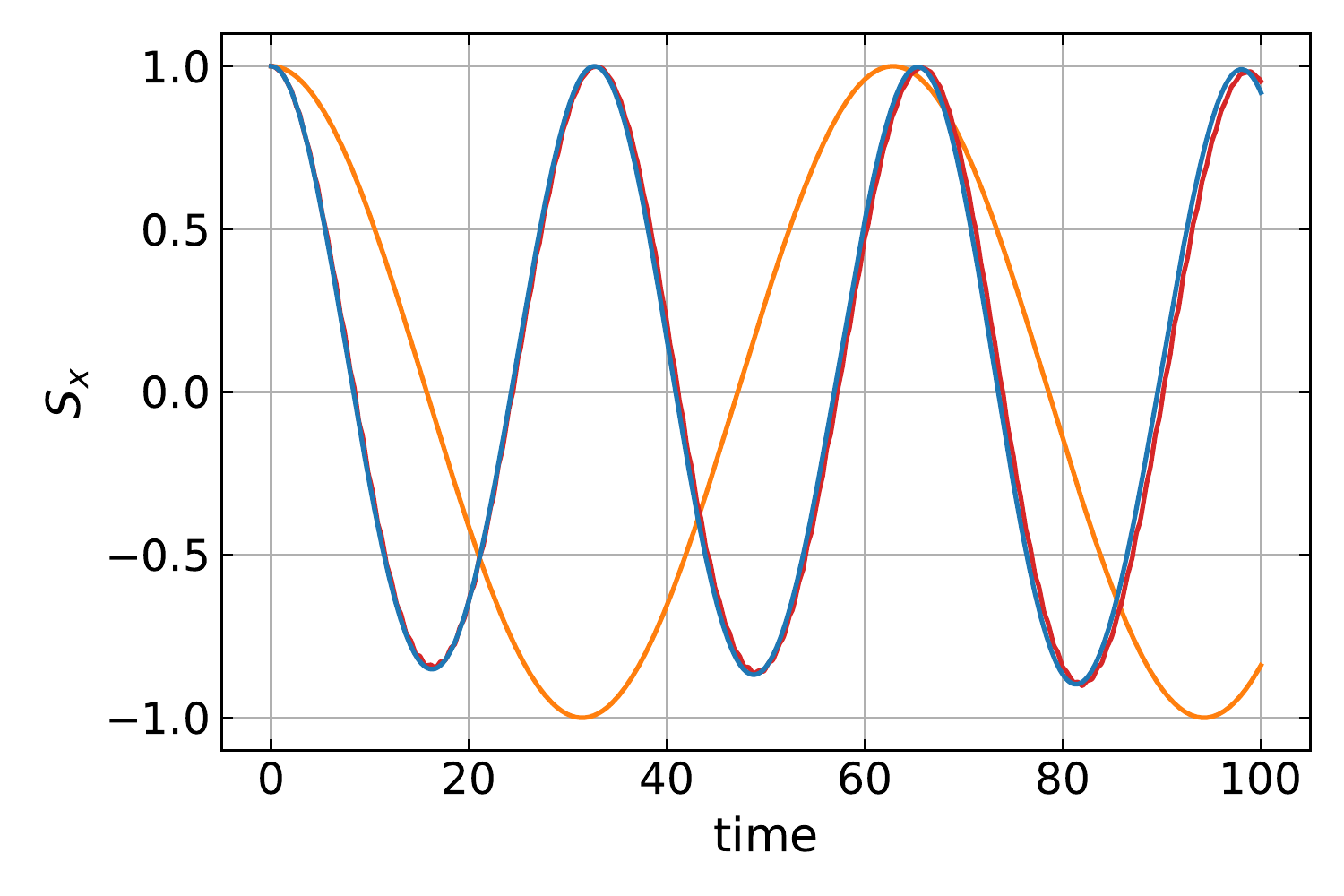}
\caption{
Time dependence of the $x$-component of the classical spin for $L=10$, $J=15$, $T=1$, $B=0.1$.
Results as obtained from ASD ($n=1$, orange), NA-SD with $n=2$ (blue), and the full theory (red).
}
\label{fig:fullad}
\end{figure}

Numerical results for a strong coupling $J=15$ are displayed in Fig.\ \ref{fig:fullad}. 
The full theory (see red curve) does predict an oscillatory motion of $S_{x}$ as expected for precessional dynamics. 
However, the precession is not perfect:
Note, e.g., that $S_{x}$ does not reach its minimum value $S_{x}=-1$, while $S_{x} \approx +1$ after a full period. 
In fact, the precession does not take place in the $x$-$y$ plane but within a plane that is a somewhat tilted and, furthermore, the plane normal $\boldsymbol{n} \propto \ff S \times \dot{\ff S}$ is slightly time dependent.

The most important effect seen in Fig.\ \ref{fig:fullad}, however, is the strongly enhanced precession frequency 
$\omega_{\rm p} \approx 0.19$, which is close to {\em twice} the Larmor freqency $\omega_{\rm L} = B = 0.1$.
This anomalous precession frequency $\omega_{\rm p}$ is clearly at variance with the naive expectation and must therefore result from the renormalization factor $1 / ( 1 - \boldsymbol{S} \langle \boldsymbol{\Omega} \rangle )$ in \refeq{ren}. 
In fact, the full theory (red) almost perfectly agrees with the prediction of the non-abelian spin-dynamics (NA-SD) theory (blue), when spanning the low-energy subspace $\ca E_{n} (\{ \ff S\})$ by the instantaneous ground and first excited state, i.e., for $n=2$.

\begin{figure}[t]
\centering
\includegraphics[width=0.95\columnwidth]{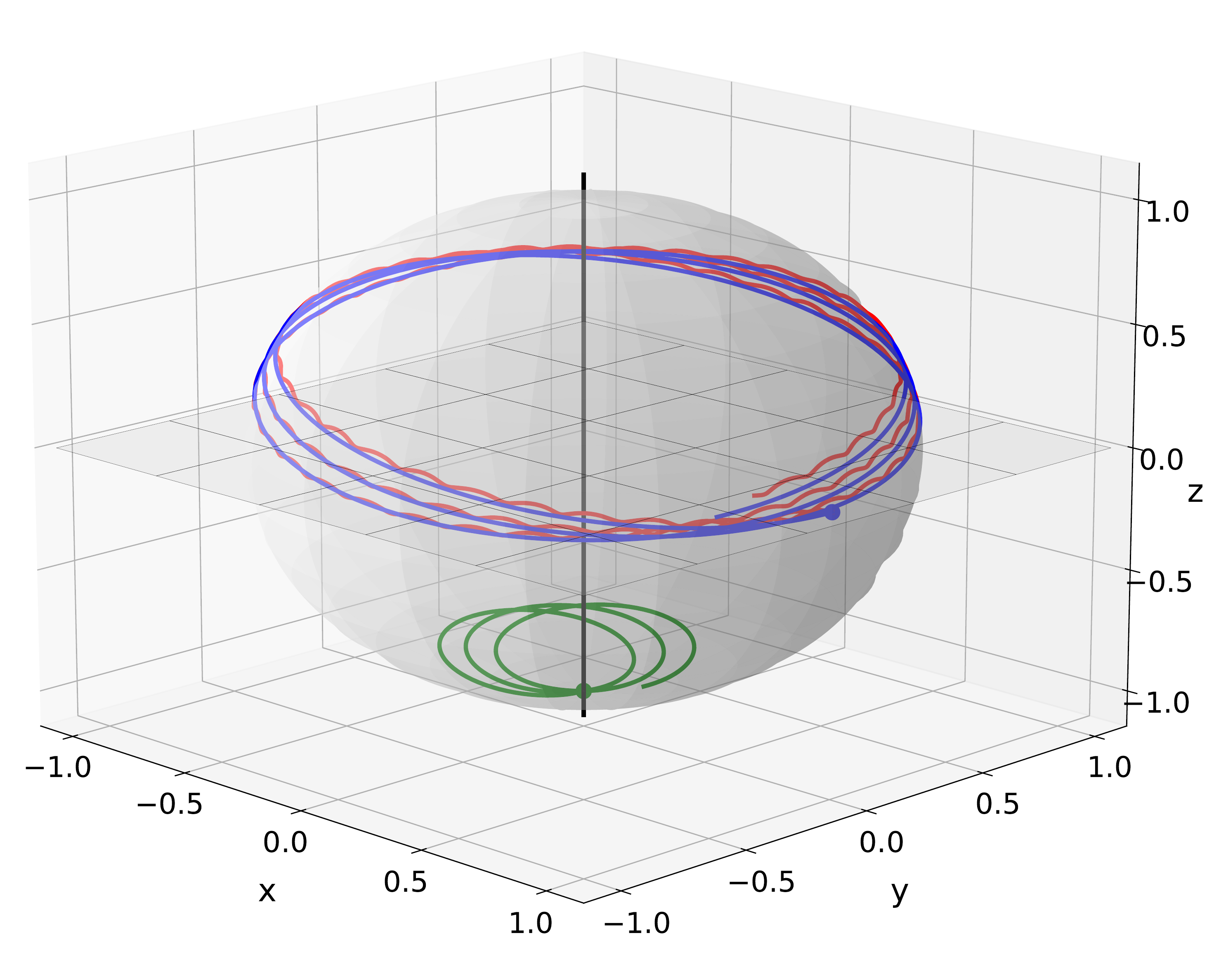}
\caption{
The same as in Fig.\ \ref{fig:fullad} for the NA-SD (blue curve) and the full theory (red) but displayed on a classical Bloch sphere. 
The blue dot marks the spin position at time $t=0$. 
Green curve: unit vector $\ff n$ normal to the instantaneous precession plane.
The trajectories are shown for $0 \le t \le 100$.
}
\label{fig:3d}
\end{figure}

Fig.\ \ref{fig:3d} presents the same results of the NA-SD (blue curve) and the full theory (red) in a classical-Bloch-sphere representation. 
At $t=0$, the motion of $\ff S$ starts at $\ff S = (1,0,0)$ (see blue dot) and completes about three full periods up to the maximum propagation time $t=100$. 
The dynamics is close to a planar precession but the instantaneous plane normal $\ff n$ (green curve) exhibits a weak time dependence and precesses itself around an axis that is somewhat tilted against the $z$ axis. 
The full theory exhibits some additional wiggles which can also be seen in Fig.\ \ref{fig:fullad} already and which are absent in the NA-SD. 
A low-energy subspace with more than $n=2$ dimensions would be necessary to capture this effect. 
Apart from that, however, there is an almost perfect agreement of the NA-SD results with the results of the full theory. 

While this is very satisfying and underpins the construction of the NA-SD, there is an interesting problem remaining: 
Comparing with the $n=1$ theory, i.e., with ASD, there is strong discrepancy. 
ASD (see orange curve in Fig.\ \ref{fig:fullad}) does in fact yield the same result as the naive adiabatic picture for the present setup since the ($n=1$) spin-Berry curvature vanishes identically: $\ff \Omega=0$.
This has been noted in Ref.\ \cite{SP17} already, and the anomalous precession frequency has been explained by referring to an effective two-spin model $H_{\rm two-spin} = J\ff s_{i_{0}}\ff S - \ff B \ff S$ which disregards the presence of the sites $i \ne i_{0}$, which can be argued to the justified in the strong-$J$ regime.
The two-spin model indeed predicts $\ff \Omega = \frac12 \ff S$, so that the renormalization factor 
$1/(1-\ff \Omega \ff S) = 2$, which is in reasonable agreement with the results of the full theory. 

The remaining problem is to clarify why, for the full model (\ref{eq:min}), the $n=1$ spin-Berry curvature vanishes. 
One should note that there is actually an odd-even effect. 
For an odd number of sites $L$, the spin-Berry curvature is in fact finite, and the agreement with the full theory is satisfying already at the $n=1$ level, while extending the effective theory to $n=2$ yields smaller corrections only. 

The odd-even effect can in fact be explained by a combination of time-reversal symmetry and the fact that a local spin-dependent perturbation applied to a non-magnetic ground state cannot induced a finite spin polarization in one dimension.
For $J = 0$, the ground state $| \Psi_{0} \rangle$ of an spin-SU(2)-symmetric tight-binding model is a total-spin singlet. 
For $J > 0$, we have $| \Psi_{0} \rangle = | \Psi_{0} (\ff S) \rangle$, where the $\ff S$ dependence is induced by the local perturbation $J \ff S \ff s_{i_{0}}$. 
Assuming, without loss of generality, that $\ff S = S \ff e_{z}$, it is given by a Slater determinant of the form 
$| \Psi_{0} (\ff S) \rangle = \prod_{k} \prod_{k'} c^{\dagger}_{k\uparrow} c^{\dagger}_{k'\downarrow} | \mbox{vac.} \rangle$, where $k,k'$ refer to the occupied spin-$\uparrow$ and spin-$\downarrow$ eigenstates of the full Hamiltonian, including the perturbation, with eigenenergies $\varepsilon_{\uparrow}(k)$ and $\varepsilon_{\downarrow}(k)$, respectively. 

For a one-dimensional particle-hole symmetric tight-binding model at half-filling, a local spin-dependent but spin-diagonal perturbation $J S_{z} s_{i_{0}z}$ does not change the number of $\uparrow$ and of $\downarrow$ eigenstates with eigenenergies $\varepsilon_{\uparrow}(k) , \varepsilon_{\downarrow}(k) <0$, for arbitrary coupling strength $J$ \cite{KST99}.
This implies that for even $L$ and at half-filling $N=L$, we must have $N_{\uparrow}= N_{\downarrow}= N/2$. 
Consequently, the number of factors in the Slater determinant, labelled by $k$ and $k'$, is the same, and thus $| \Psi_{0} (\ff S) \rangle $ is still a total-spin singlet (constructed from $\ff S$-dependent one-particle states), irrespective of the strength of the perturbation $J$.
This argument holds for any direction of $\ff S$ and thus implies that $\Theta | \Psi_{0} (\ff S) \rangle = | \Psi_{0} (\ff S) \rangle$, i.e., the ground state is invariant under time reversal $\Theta$ for all $\ff S$.
Hence, the same holds for its $\ff S$ derivative: $\Theta | \partial_{\ff S} \Psi_{0} (\ff S) \rangle = | \partial_{\ff S} \Psi_{0} (\ff S) \rangle$.
Some details on the invariance under time reversal are given in Appendix \ref{sec:trs}.

Specializing \refeq{sbcq} to the adiabatic case $n=1$ we thus have 
$\ff \Omega = \frac12 \sum_{\alpha \beta\gamma} \varepsilon_{\alpha\beta\gamma} \ff e_{\alpha}\Omega_{\beta\gamma}$ with
\ba
\Omega_{\beta\gamma} 
&=& 
i \left[
\bra{\partial_{S_{\beta}} \Psi_{0}} \ket{\partial_{S_{\gamma}}\Psi_{0}} 
- 
(\beta  \leftrightarrow \gamma)
\right]
\nonumber \\
&=&
- 2 \,  \mbox{Im}
\bra{\partial_{S_{\beta}} \Psi_{0}} \ket{\partial_{S_{\gamma}}\Psi_{0}} 
\nonumber \\
&=&
- 2 \,  \mbox{Im}
\bra{\partial_{S_{\beta}} \Psi_{0}} \Theta^{\dagger} \Theta \ket{\partial_{S_{\gamma}}\Psi_{0}}^{\ast}
\nonumber \\
&=&
- 2 \,  \mbox{Im}
\bra{\partial_{S_{\beta}} \Psi_{0}} \ket{\partial_{S_{\gamma}}\Psi_{0}}^{\ast}
=0 \:, 
\ea
where we have exploited the anti-unitarity of $\Theta$.
In an extension of the discussion of Sec.\ \ref{sec:tr} for the weak-$J$ case, we can thus infer that the abelian  spin-Berry curvature must vanish for even $L$ and arbitrary $J$ in one dimension.
Let us emphasize that the argument cannot be transferred to the non-abelian case.
For $n>1$, we have $\langle \ff \Omega \rangle \ne 0$ in general.

\subsection{Nutation}

Apart from the precessional motion, the classical-spin dynamics also exhibits nutational oscillations with a frequency that is in general different from the precession frequency. 
The nutation is most easily seen in an oscillatory behavior of the $z$ component of the classical spin: 
The field points into the $z$ direction, $\ff B = B \ff e_{z}$, such that the $z$ component of the torque on $\ff S$ due to the field must vanish $(\ff B \times \ff S)_z = 0$. 
A nonzero time derivative $\dot{S}_z \ne 0$ is, therefore, solely due to the exchange coupling and directly proportional to $J (\langle \ff s_{i_0} \rangle \times \ff S)_z$.  

As such, a nutational motion cannot be captured by $n=1$ adiabatic spin-dynamics (ASD) theory: 
The adiabatic constraint and a simple symmetry argument immediately imply that the ground-state local moment $\langle \ff s_{i_0} \rangle = \bra{\Psi_{0}(\ff S)} \ff s_{i_0} \ket{\Psi_{0}(\ff S)}$ must be strictly antiparallel (for $J>0$) to $\ff S$, which in turn  implies that $S_{z}$ is a constant of motion.
The adiabatic spin dynamics is thus perfectly precessional albeit, opposed to naive adiabatic theory, with a renormalized precession frequency, as already discussed above.

\begin{figure}[t]
\centering
\includegraphics[width=0.95\columnwidth]{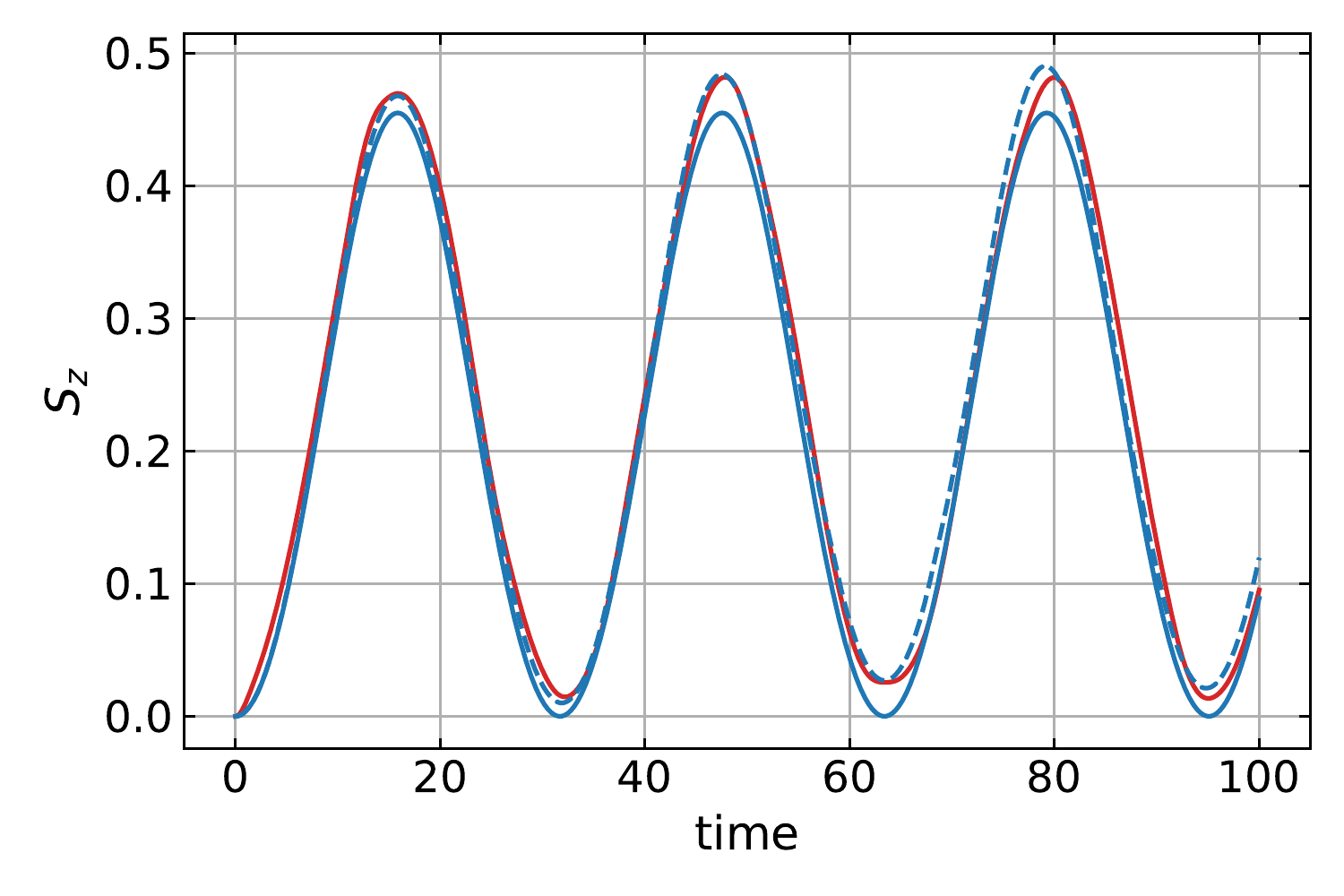}
\caption{
Time dependence of the $z$-component of the classical spin as obtained from $n=2$ (solid blue curve) and from $n=4$ (dashed blue curve) NA-SD, compared to the result (red curve) of the full theory. $L = 11$, $J=1$, $B = 0.1$.
}
\label{fig:ndep}
\end{figure}

Numerical results for $L=11$ as obtained from non-abelian spin-dynamics theory with $n=2$, see Fig.\ \ref{fig:ndep} (blue curve), show that there can be a considerable variation of the amplitude of the $z$ component of $\ff S$. 
The nutational oscillation is perfectly harmonic and $S_{z}$ stays non-negative, when starting with $S_{z}=0$ at $t=0$.
As compared with the $S_{z}$ dynamics predicted by the full theory (red curve), the step from $n=1$ (ASD) to $n=2$ (most simple variant of NA-SD) is in fact the essential one, and the results for $n=2$ are already close to those of the full theory. 
The latter, however, predicts a slight deviation from perfectly harmonic nutational motion, which is not reproduced with $n=2$ but can be captured with an improved ($n=4$) approximation within the NA-SD (dashed blue curve).
A further increase of $n$ becomes technically more and more involved and has also been found to improve the results in a non-monotonic way only. 
It is thus very fortunate that the main improvement of the $n=1$ ASD is already achieved with $n=2$ NA-SD.
For the rest of the discussion, we will therefore stick to the $n=2$ case.

\begin{figure}[tb]
\centering
\includegraphics[width=0.95\columnwidth]{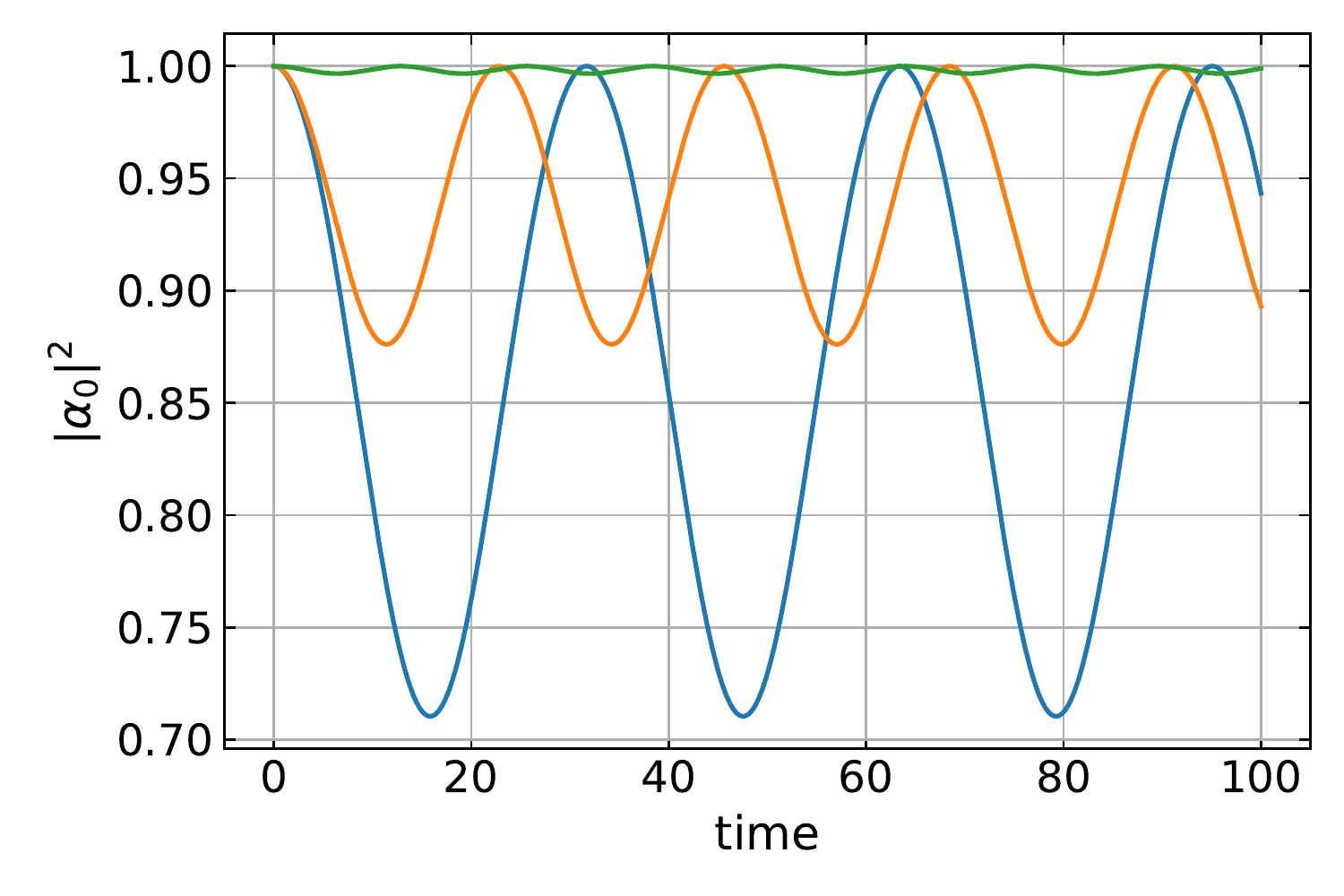}
\caption{
Time dependence of the ground-state weight $|\alpha_{0}|^{2}$ as obtained from NA-SD with $n=2$ for a system with $L=11$, for $B=0.1$, and for various coupling strengths $J=1$ (blue), $J=2$ (orange), and $J=10$ (green).
}
\label{fig:alpha11}
\end{figure}

The physical cause of the nutation can be traced back to the time-dependent admixture of the first excited state $|\Psi_{1}(\ff S(t)) \rangle$ to the instantaneous ground state $|\Psi_{0}(\ff S(t))\rangle$. 
Fig.\ \ref{fig:alpha11} for $J=1$ (blue curve) displays the absolute square of the ground-state coefficient $|\alpha_{0}|^{2}$ as function of propagation time corresponding to the $n=2$ NA-SD result for $S_{z}$ in Fig.\ \ref{fig:ndep}.
Note that we have $|\alpha_{1}|^{2} = 1 - |\alpha_{0}|^{2}$ for $n=2$. 
As for $t=0$ the conduction-electron system is prepared as the ground state of $\hat{H}$, the ground-state weight $|\alpha_{0}|^{2} = 1$ initially. 
In the course of time, there is a weight transfer to the first excited state, which results in a significant reduction of the ground-state weight down to a minimal value of $|\alpha_{0}|^{2} \approx 0.72$.
Within the $n=2$ NA-SD, the time-dependent weight transfer is perfectly harmonic, and its frequency is exactly the same as the nutation frequency of $S_{z}$ (see Fig.\ \ref{fig:ndep}). 

\begin{figure}[b]
\centering
\includegraphics[width=0.95\columnwidth]{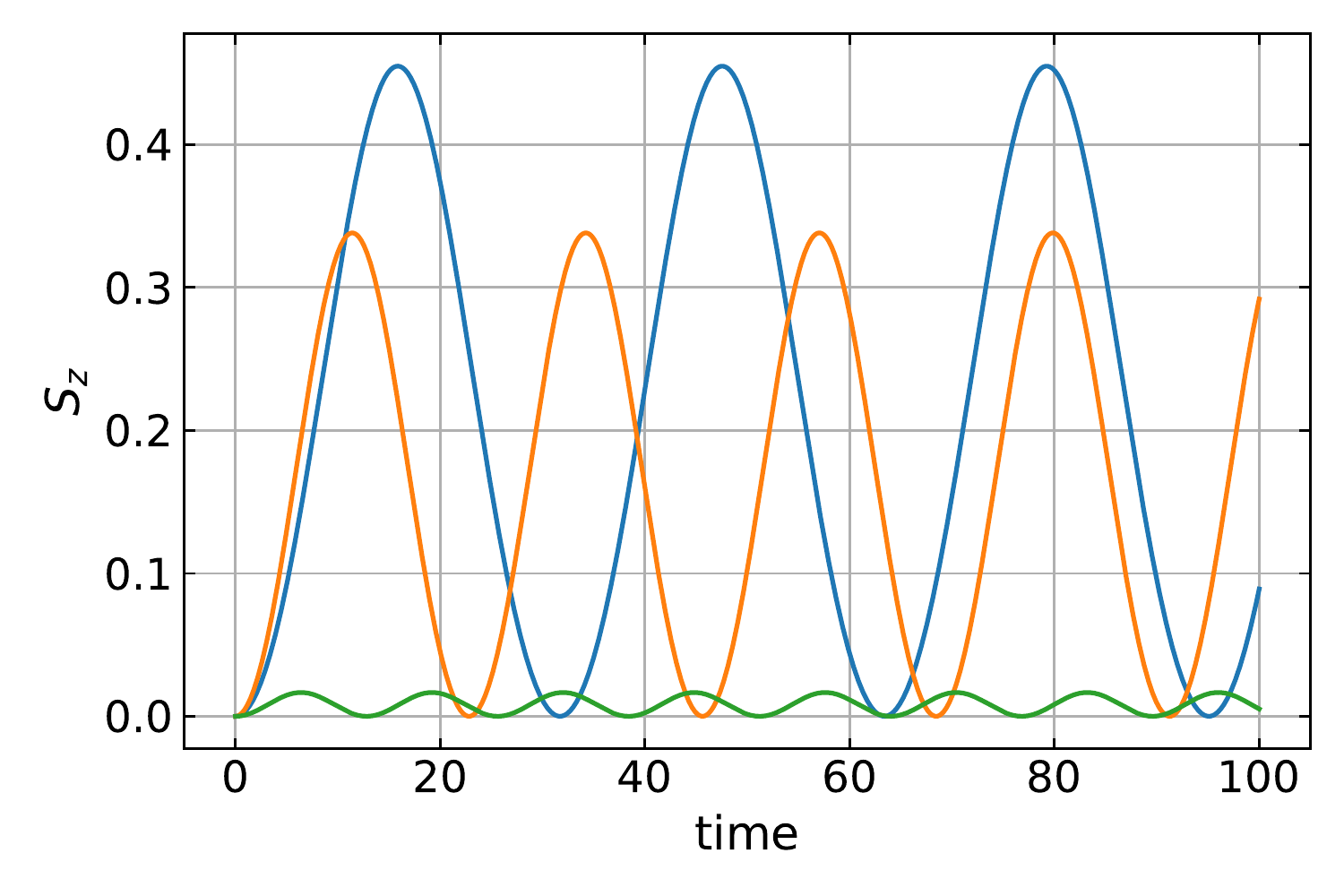}
\caption{
The same as Fig.\ \ref{fig:alpha11} but for the time dependence of $S_{z}$.
Coupling strengths: $J=1$ (blue), $J=2$ (orange), and $J=10$ (green).
}
\label{fig:sz11}
\end{figure}

\begin{figure}[tb]
	\centering
	\includegraphics[width=0.95\columnwidth]{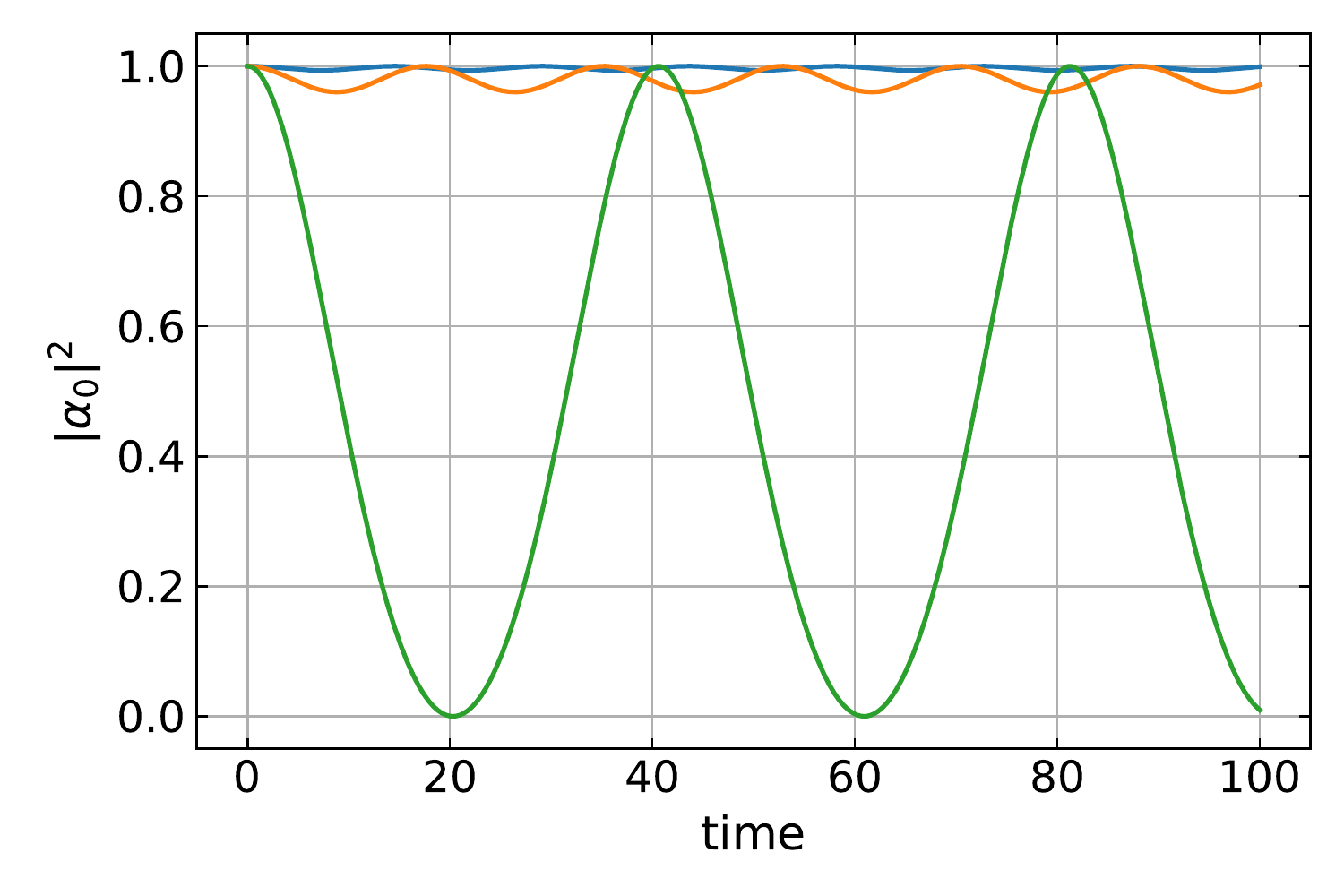}
\caption{
The same as Fig.\ \ref{fig:alpha11} but for $L=10$.
Note that the same color coding is used. 
Coupling strengths: $J=1$ (blue), $J=2$ (orange), and $J=10$ (green).
}
\label{fig:alpha10}
\end{figure}

\begin{figure}[b]
\centering
\includegraphics[width=0.95\columnwidth]{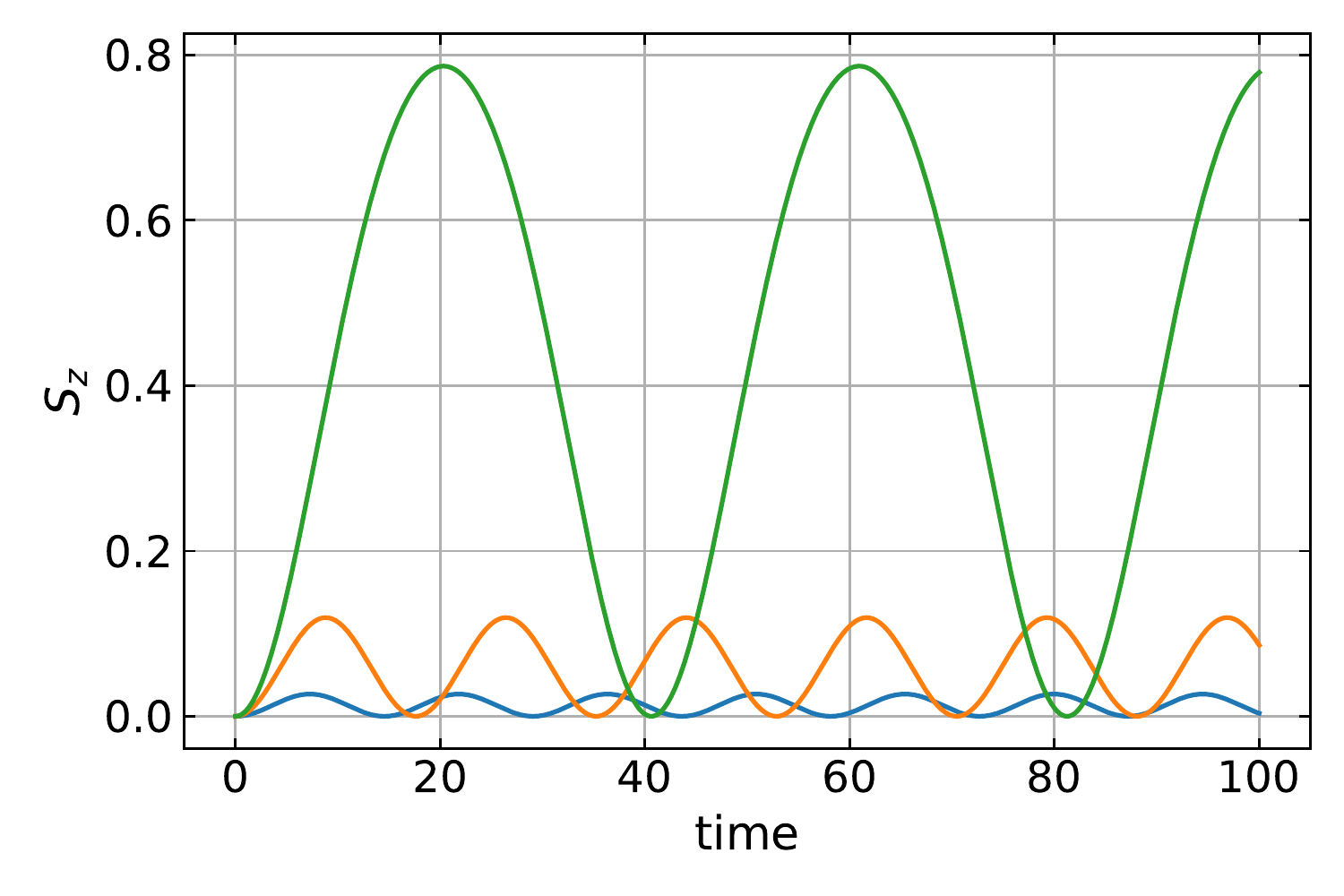}
\caption{
The same as Fig.\ \ref{fig:alpha10} but for the time dependence of $S_{z}$.
Coupling strengths: $J=1$ (blue), $J=2$ (orange), and $J=10$ (green).
}
\label{fig:sz10}
\end{figure}

Increasing the coupling strength $J$, results in a weaker admixture of the first excited state, as can be seen by the results for $J=2$ (orange) and $J=10$ (green) in Fig.\ \ref{fig:alpha11}.
This is accompanied by an increasing frequency of the time-dependent weight transfer. 
Again, this frequency is precisely the nutation frequency that is observed in the time dependence of $S_{z}$, which is displayed in Fig.\ \ref{fig:sz11} for the different coupling strengths. 
We also note that this is unrelated with the precession frequency which is much less $J$ dependent.
Furthermore, also the $J$ dependence of the minimal (maximal) amplitude shows the same trend for both, the ground-state weight and for $S_{z}$, respectively.

Compared to the standard perturbative linear-response approach discussed in the introduction, our approach thus provides an alternative explanation of nutational spin dynamics. 
As in the standard theory, nutation is the first phenomenon that is found in a systematic expansion starting around the adiabatic limit, namely Taylor expansion in the retardation time on the one hand and expansion in the dimension of the instantaneous low-energy subspace on the other.
Another important difference is that the NA-SD is formulated for a closed system while the standard theory relies on a formalism for open quantum systems. 
This also explains that the standard approach necessarily predicts non-conserving Gilbert damping accompanying the nutation motion.

\begin{figure}[t]
\centering
\includegraphics[width=0.95\columnwidth]{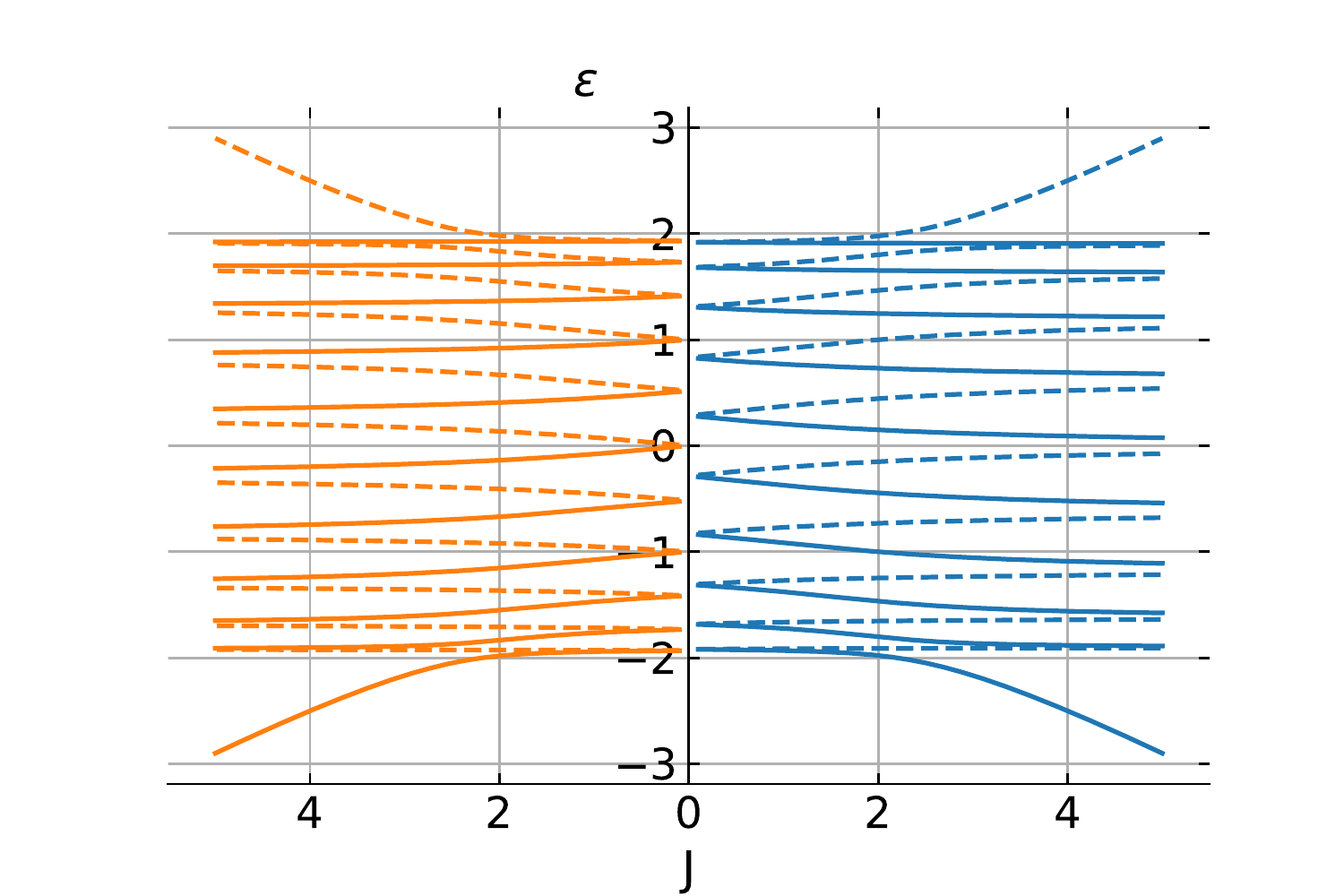}
\caption{
$J$ dependence of the single-particle eigenenergies for $L=10$ (right) and $L=11$ (left). 
$B = 0.1$. 
}
\label{fig:jdep}
\end{figure}

\begin{figure}[b]
	\centering
	\includegraphics[width=0.95\columnwidth]{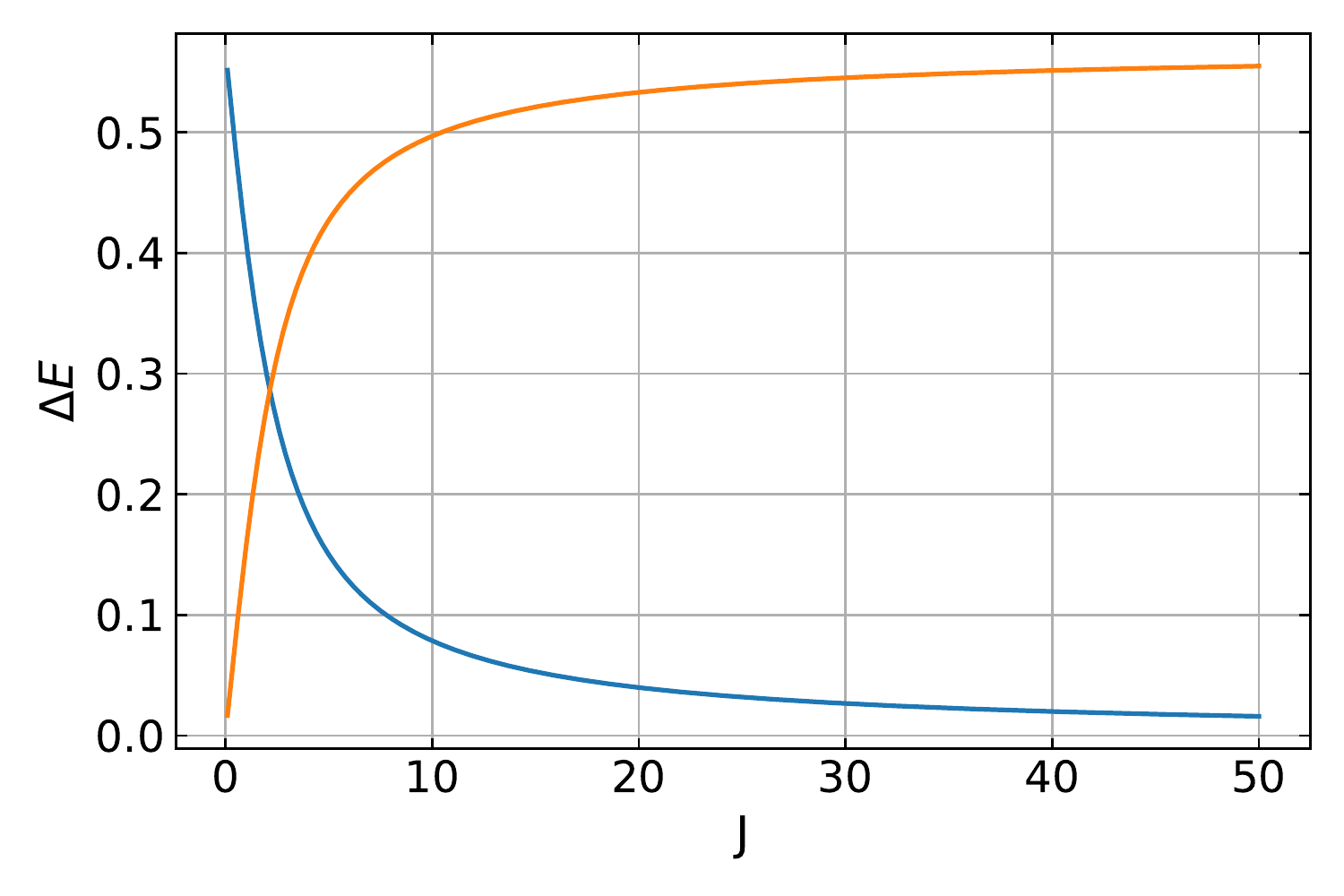}
\caption{
The finite-size energy gap $\Delta E$ between the ground state and the first excited state as function of $J$ for $L=10$ (blue) and $L=11$ (orange).
$B=0.1$.
}
\label{fig:gap}
\end{figure}

The time dependence of the weight $|\alpha_{0}|^{2}$, as shown in Figs.\ \ref{fig:alpha11} and {\ref{fig:alpha10}, is reminiscent of the Rabi oscillations of the ground-state occupation in a simple two-level system driven by an oscillatory time-dependent external field. 
In our case the driving is due to the classical spin which is precessing around the axis of the magnetic field.
However, the case is more complicated.
Opposed to the standard Rabi setup \cite{LeB06}, the ``two-level system'' emerging in the ($n=2$) NA-SD is itself time-dependent, has a feedback on the classical spin induced by the spin-Berry curvature via the geometrical torque, and $\ff S$ couples locally rather than globally to a time-dependent and in general only partially polarized local magnetic moment. 

Let us return to the results for the ground-state weight for $L=11$ shown in Fig.\ \ref{fig:alpha11}.
It is tempting to interpret the decrease of the amplitude of the oscillations of $|\alpha_{0}|^{2}$ with increasing $J$ as a consequence of approaching the adiabatic limit, where $|\alpha_{0}|^{2} =1$.
In fact, this trend is consistent with the time-averaged angle enclosed by $\ff S$ and $\langle \ff s_{i_{0}} \rangle$ approaching $180^{\circ}$ with increasing $J$ (see Fig.\ \ref{fig:angle}). 
However, for a tight-binding chain with an even number sites ($L=10$), see the data in Fig.\ \ref{fig:alpha10}, we find that the oscillation amplitude of $|\alpha_{0}|^{2}$ grows with increasing $J$.
We conclude that there is an odd-even effect not only with respect to the precessional but also to the nutational dynamics.

For an explanation of the effect, we consider the $2L$ single-particle eigenenergies $\varepsilon_{k}$ of the minimal model \refeq{min}. 
Their $J$ dependence is shown in Fig.\ \ref{fig:jdep} for $L=10$ (right, blue lines) and $L=11$ (left, orange lines).
Only at $J=0$ are the eigenenergies spin-degenerate, any finite $J>0$ immediately lifts this degeneracy.
Consistent with analytical results available for tridiagonal pseudo-Toeplitz matrices \cite{KST99}, we find that the $\varepsilon_{k}(J)$ curves do not intersect and that a finite ``critical'' coupling $J \approx 2$ is necessary to split off a pair of bound states,  localized in the vicinity of $i_{0}$, from the ``continuum'' of delocalized states. 
Importantly, however, we note that the finite-size gap $\Delta E$ between the highest occupied and the lowest unoccupied eigenenergy, right below and right above $\varepsilon=0$, respectively, shows opposite trends for $L=10$ and $L=11$. 

The $J$ dependence of the gap is displayed in Fig.\ \ref{fig:gap}. 
We note that $\Delta E$ monotonically shrinks with $J$ for $L=10$ (blue lines) and grows with $J$ for $L=11$. 
This is also characteristic in general, for systems with an even and odd number of sites, respectively. 
According to the adiabatic theorem \cite{LeB06}, the real-time dynamics is close to adiabatic if the gap size is large compared to the inverse $\tau^{-1}$ of the typical time scale $\tau$. 
Here, this can be estimated as given by $\tau^{-1} \sim B = 0.1$. 
For the case $L=11$, this indeed implies that the adiabatic limit is approached with increasing $J$, while for $L=10$ a decreasing $J$ favours adiabatic dynamics. 
This also explains the different $J$ dependence of the amplitudes of the nutational oscillations of $S_{z}$ shown in Figs.\ \ref{fig:sz11} and \ref{fig:sz10}, respectively.

\section{Concluding discussion}
\label{sec:con}

Systems of a single or a few quantum spins coupled to an extended lattice fermion model pose notoriously difficult quantum many-body problems. 
Here, by treating the impurity spins as classical objects with a dynamics that is slow as compared to the typical electronic time scales, we have concentrated on a simplified case with the ambition to exactly trace out the high-energy scales and to arrive at an effective low-energy theory that, apart from the classical spins, includes a minimal number of electronic degrees of freedom. 
Our approach in fact represents a systematic extension of the previously proposed adiabatic spin dynamics (ASD) theory \cite{SP17}, where unconventional spin dynamics was observed to result from a geometrical spin torque.

For systems where the typical spin-dynamics time scale is much slower than the time scale of the electron dynamics, the adiabatic theorem, in case of gapped systems, tells us that the electron state at an instant of time $t$ is the ground state of the electronic Hamiltonian for the given spin configuration at $t$. 
Alternatively and more general, one may argue that adiabatic dynamics is due to fast electronic relaxation processes dissipating the excess energy to the bulk of the system or to external baths. 
These standard arguments and more explicit criteria, which typically motivate a purely adiabatic theory, are rarely controllable and hardly ever fully met in applications to realistic systems.
In most practical cases, it is a priori extremely difficult to decide whether or not the dynamics is adiabatic.
Our approach therefore aims at a straightforward way to improve the adiabatic spin-dynamics theory in an, at least in principle,  systematic manner.

As the central and sole approximation we assume that the electronic state at any instant of time $t$ lies in the $n$-dimensional low-energy sector spanned by the instantaneous ground state, realized for the classical-spin configuration at time $t$, and the corresponding lowest $n-1$ instantaneous excited states of the electron system.
The approximation is implemented as a holonomic constraint within a Lagrange formalism. 
We have seen that the effective low-energy theory unfolds itself straightforwardly and naturally takes the form of a non-abelian gauge theory, where the non-abelian spin-Berry connection and spin-Berry curvature enter the resulting effective equations of motions for the electronic state and for the spins.
The gauge freedom is given by the arbitrary choice of an orthonormal basis in the instantaneous low-energy subspace of the electron system.
SU(n) gauge transformations leave observables invariant. 
The number $n$ of states considered in the non-abelian spin dynamics (NA-SD) theory can be seen as a control parameter, so that comparing results for different $n$ allows us to check the validity of the approach, at least in principle.

The physically interesting point of the emergent low-energy theory is that the spin dynamics is crucially affected by the gauge-invariant expectation value of the (gauge-covariant) spin-Berry curvature, i.e., by an additional geometrical spin torque. 
In the ASD ($n=1$) a non-zero spin-Berry curvature is obtained for systems with broken time-reversal symmetry only. 
Opposed to ASD ($n=1$), however, the non-abelian spin dynamics (NA-SD) theory incorporates a spin-Berry curvature tensor, the elements of which are generically non-zero even in the more common time-reversal-symmetric case and both, for the anti-unitary time-reversal operator squaring to $+1$ and to $-1$. 
The NA-SD formalism also provides an elegant and straightforward explanation for the odd-even effect observed as function of the system size in the simpler ASD \cite{SP17}.

Applications of the NA-SD theory are promising in cases, where (i) the classical-spin approximation is reasonable, e.g., for magnetic atoms with high spin quantum numbers or, more generally, with well-developed local magnetic moments, which are stable on time scales exceeding all other relevant time scales of the full system. 
This excludes, e.g., Kondo systems with a fast screening of the local moment.
Strong magnetic anisotropies at surfaces or interfaces, on the other hand, can favor extremely stable magnetic moments with respect to both, longitudinal and transversal spin fluctuations \cite{Wie09}.

(ii) As regards the electron system, the amount of energy pumped in with the initial excitation must be small compared to the lowest electron excitation energies, such that a low-dimensional instantaneous low-energy subspace can fully capture the essential dynamics.
Such situations could be realized in case of magnetic atoms coupled to tight-binding systems with essentially a finite number of orbitals, e.g., to metallic nanoislands supported by an insulating substrate \cite{Wie09} or in nanowires \cite{BN20}, for example. 
Correlated molecular magnetic systems are interesting as well, particularly in cases with a degenerate ground-state manifold (see Ref.\ \cite{RPPK22} for an instructive example), which naturally defines the low-energy subspace.
In case of formally infinite, e.g., condensed-matter systems, NA-SD may be applicable whenever there is a low-energy sector with a finite gap to excited states at higher energies, such as insulating systems with a symmetry-induced degenerate ground state. 
Topological insulators with gapless edge modes, e.g., Chern or Z$_{2}$ insulators, represent another class of systems which are worth to be considered, and the study of the relation between different Berry curvatures, the spin-Berry curvature considered here and the conventional Berry curvature of topological band theory is expected to be particularly instructive.
The real-time dynamics of classical spins coupled to the edge of a one-dimensional spinful Su-Schrieffer-Heger model \cite{EP21} and to a two-dimensional spinful Kane-Mele model \cite{QP22} have been discussed recently.
In the former case, the low-energy subspace (at one edge) is spanned by two quantum states only. 
For the Z$_{2}$ Kane-Mele nanoribbon, the helical edge modes form a continuum but with an extremely small phase space for spin excitations, which suggests that considering a finite number of basis states for the low-energy sector could be a reasonably good approximation.

For classical spins coupled to gapless metallic bulk systems, any low-energy sector is formally infinite-dimensional. 
While the adiabatic theorem does not apply to this case, one still expects that a low-energy subspace defined by a certain maximum excitation energy $\Delta E$ above the many-electron ground state could reliably capture the electron dynamics, depending on the initial excitation energy pumped into the system.
If the electron system may be treated in the independent-electron approximation, the application of NA-SD is well conceivable, since it merely involves diagonalization of the single-electron hopping matrix and computation of matrix elements of two-electron operators with two-electron and two-hole excited states above the Fermi sea (see Appendix \ref{sec:comp}).
By varying $\Delta E$, the reliability of the approximation can be tested.

Here, as a proof of principle, we performed numerical calculations for a minimal but non-trivial model consisting of a single impurity spin coupled to the first site of a one-dimensional non-interacting tight-binding model with a small number of $L$ sites.
The real-time dynamics is initiated by a sudden change of the direction of a local magnetic field coupled to the impurity spin only. 
Results obtained from ASD ($n=1$) and NA-SD (for $n=2$ and $n=4$) have been checked against results obtained from the numerical solution of the full, unconstrained set of equations of motion for the coupled spin-electron system.
We find that the NA-SD reproduces the anomalous precession frequency that is already predicted by ASD for systems with an odd number of sites $L$. 
For even $L$, NA-SD correctly predicts anomalous precession, which is absent in the purely adiabatic approach. 
This deficiency of the ASD can be explained by a symmetry analysis.
Depending on the coupling strength $J$, the dynamics of the impurity spin can exhibit a considerable nutational motion. 
As judged by comparison with the full theory, this more subtle effect is almost quantitatively covered with NA-SA for $n=2$. 
NA-SD calculations for $n=4$ show an even closer agreement with the full theory. 
\\

\acknowledgments
We acknowledge support from the Deutsche Forschungsgemeinschaft (DFG, German Science Foundation) through FOR 5249-449872909 (Project P8) and the European Research Council via Synergy Grant 854843-FASTCORR.

\appendix

\begin{widetext}
	
\section{Explicit form of the equation of motion for the classical spins}
\label{sec:exp}

The equations of motion \refeq{eomss} for the classical spins derived in Sec.\ \ref{sec:eom} are implicit differential equations. 
An explicit form, however, is more convenient for the numerical evaluation. 
Here, we briefly discuss a corresponding reformulation. 
We start with \refeq{eomss} and apply $\times \boldsymbol{S}_m$ from the right. 
This yields
\be
\dot{\boldsymbol{S}}_m 
=
\langle \partial_{\ff S_{m}} \hat{H}_{\rm int} \rangle \times \boldsymbol{S}_m
+ 
\partial_{\ff S_{m}} H_{\rm cl} \times \boldsymbol{S}_m 
+ 
\sum_\delta \sum_{k\gamma} 
\left( \sum_{\alpha\beta} \varepsilon_{\alpha \beta \delta} S_{m\beta} \langle {\Omega} \rangle_{k\gamma, m\alpha} \right) 
\dot{S}_{k\gamma} \boldsymbol{e}_\delta 
\: .
\labeq{exp1}
\ee
Next, we combine the components of all $M$ spins in a single $3M$-dimensional column, 
\be
\boldsymbol{\mathcal{S}} := (\boldsymbol{S}_1, \boldsymbol{S}_2, \dots)^T = \sum_{m=1}^{M} \boldsymbol{e}^{M}_{m} \otimes \boldsymbol{S}_m \: .
\ee
Here $\boldsymbol{e}^{M}_{m}$ is the $m$-th canonical $M$-dimensional unit vector and $\otimes$ denotes the Kronecker product. 
Writing 
$\chi_{m\delta,k\gamma} =  \sum_{\alpha\beta} \varepsilon_{\alpha\beta\delta} S_{m\beta} \langle {\Omega} \rangle_{k\gamma, m\alpha}$ for short, the last term on the right-hand side of \refeq{exp1} can be written as
$\sum_\delta ( \underline{\chi} \dot{\boldsymbol{\mathcal{S}}} )_{m\delta} \boldsymbol{e}_\delta$, and we find the explicit form of the $3M$-dimensional system of differential equations of motion:
\be
\dot{\boldsymbol{\mathcal{S}}} 
= 
\Big(\mathbb{1}  - \underline{\chi}\Big)^{-1} \cdot
\left(
\sum_{m} \boldsymbol{e}^{M}_{m} \otimes \left( 
\langle \partial_{\ff S_{m}} \hat{H}_{\rm int} \rangle \times \boldsymbol{S}_m 
+ 
\partial_{\ff S_{m}} H_{\rm cl} \times \boldsymbol{S}_m 
\right) 
\right) 
\: .
\ee
This involves an inversion of the $3M$-dimensional matrix $\mathbb{1} - \underline{\chi}$.

\section{Normalisation conditions}
\label{sec:norm}

The equations of motion \refeq{eomss} and \refeq{wfs} respect the normalisation conditions \refeq{ncon}. 
We start with the wave-function normalization. 
\refeq{wfs} implies 
\be
i
\sum_{i}\alpha^{\ast}_{i}(\partial_{t}\alpha_i)
=
 \sum_{ij}\alpha^{\ast}_{i}\alpha_{j}\bra{\Psi_{i}}\hat{H}\ket{\Psi_{j}}
- i \sum_{ij} \alpha^{\ast}_{i} \alpha_{j} \bra{\Psi_{i}} \partial_{t} \ket{\Psi_{j}}
=
- i \sum_{i}(\partial_{t}\alpha^{\ast}_{i})\alpha_{i}
\: .
\ee
This yields $\partial_{t} \sum_{i} | \alpha_{i} |^{2} = 0$ as required.
Conservation of the length of the classical spins can be verified directly from their equations of motion, \refeq{eomss}, or, more conveniently by taking the scalar product of both sides of \refeq{exp1} with $\ff S_{m}$. 
This yields $\ff S_{m} \dot{\ff S}_{m} = 0$ as required.
However, conservation of the spin length has been exploited already in deriving \refeq{eomss}, directly after \refeq{eoms}.

Alternatively, we may thus explicitly take care of the normalization conditions $\boldsymbol{S}^{2}_{m}=1$ by treating them as additional constraints when deriving the equations of motion from the Lagrangian \eqref{eq:leff}. 
This is done with $M$ Lagrange multipliers $\lambda_{m}$, i.e., we replace the Lagrangian by
\be
L^{\prime}_{\text{eff}}(\{\boldsymbol{S}\}, \{\dot{\boldsymbol{S}}\},\{\alpha\},\{\alpha^{\ast}\},\{\dot{\alpha}\},\{\dot{\alpha}^{\ast}\},
\{\lambda\})=L_{\text{eff}}(\{\boldsymbol{S}\}, \{\dot{\boldsymbol{S}}\},\{\alpha\},\{\alpha^{\ast}\},\{\dot{\alpha}\},\{\dot{\alpha}^{\ast}\}) - \sum_{m} \lambda_m (\boldsymbol{S}^{2}_{m}-1)
\: , 
\ee
such that the Euler-Lagrange equation for $\lambda_m$ reads $\boldsymbol{S}^{2}_{m} = 1$.
Further, the equation of motion for a classical spin $\boldsymbol{S}_m$ is modified as
\be
0 
= 
\frac{1}{\abs{\boldsymbol{S}_m}^{3}}\dot{\boldsymbol{S}}_m\times\boldsymbol{S}_m 
+ 
\langle \partial_{\ff S_{m}} \hat{H}_{\rm int} \rangle
+
\partial_{\ff S_{m}} H_{\rm cl} 
+ 
\sum_{k} \sum_{\beta\gamma} 
\dot{S}_{k\gamma}
\langle {\Omega} \rangle_{k\gamma, m\beta}
\ff e_{\beta} 
+ 
2\lambda_m \boldsymbol{S}_m
\: .
\labeq{eommod}
\ee
Acting on both sides of the equation with $\times\boldsymbol{S}_m$ and with $\cdot\boldsymbol{S}_m$, respectively, gives a system of two equations, which is equivalent with \refeq{eommod}:
\ba
	0 &=& (\dot{\boldsymbol{S}}_m \times \boldsymbol{S}_m) \times\frac{\boldsymbol{S}_m}{\abs{\boldsymbol{S}_m}^{3}} 
	+ \langle \partial_{\ff S_{m}} \hat{H}_{\rm int} \rangle \times \boldsymbol{S}_m
	+
	\partial_{\ff S_{m}} H_{\rm cl} \times \boldsymbol{S}_m
	+ 
	\sum_{k} \sum_{\beta\gamma} 
	\dot{S}_{k\gamma}
	\langle {\Omega} \rangle_{k\gamma, m\beta}
	\ff e_{\beta} \times \boldsymbol{S}_m
\: , \nonumber \\
	0 &=& \langle \partial_{\ff S_{m}} \hat{H}_{\rm int} \rangle \cdot \boldsymbol{S}_m
	+
	\partial_{\ff S_{m}} H_{\rm cl} \cdot \boldsymbol{S}_m
	+ 
	\sum_{k} \sum_{\beta\gamma} 
	\dot{S}_{k\gamma}
	\langle {\Omega} \rangle_{k\gamma, m\beta}
	\ff e_{\beta} \cdot \boldsymbol{S}_m + 2 \lambda_m \boldsymbol{S}^{2}_{m}
\: .
\ea
Exploiting $\boldsymbol{S}^{2}_{m} = 1$ in the second equation fixes the Lagrange multipliers as
\begin{equation}
	\lambda_m = -\frac{1}{2}\left(\langle \partial_{\ff S_{m}} \hat{H}_{\rm int} \rangle \cdot \boldsymbol{S}_m
	+
	\partial_{\ff S_{m}} H_{\rm cl} \cdot \boldsymbol{S}_m
	+ 
	\sum_{k} \sum_{\beta\gamma} 
	\dot{S}_{k\gamma}
	\langle {\Omega} \rangle_{k\gamma, m\beta}
	\ff e_{\beta} \cdot \boldsymbol{S}_m\right)
\: ,
\end{equation}
while using it in the first equation reproduces the familiar equation of motion (\ref{eq:eomss}).

\section{Spin-Berry curvature in terms of a projection operator}
\label{sec:proj}

To prove \refeq{sbcq} we start from the definition \eqref{eq:curv} of the non-abelian spin-Berry curvature and insert the definition for the spin-Berry connection \refeq{con}. 
This gives
\ba
	\Omega^{(ij)}_{k\gamma, m\beta} 
	&=& 
	i \left[ \bra{\partial_{S_{k\gamma}}\Psi_{i}} \ket{\partial_{S_{m\beta}}\Psi_{j}} 
	- \bra{\partial_{S_{m\beta}}\Psi_{i}} \ket{\partial_{S_{k\gamma}}\Psi_{j}} \right] 
	\nonumber\\
	&+& 
	i \sum_{l=0}^{n-1} 
	\left[ \bra{\Psi_{i}} \partial_{S_{k\gamma}} \ket{\Psi_{l}} \bra{\Psi_{l}} \partial_{S_{m\beta}} \ket{\Psi_{j}} 
	-  
	\bra{\Psi_{i}} \partial_{S_{m\beta}} \ket{\Psi_{l}} \bra{\Psi_{l}} \partial_{S_{k\gamma}} \ket{\Psi_{j}} \right] 
	\: ,
\ea
where we have exploited the commutativity of the derivatives $\partial_{S_{k\gamma}}$ and $\partial_{S_{m\beta}}$.
Using the completeness relation and inserting a unity, 
\be
{\mathbb{1}} = \mathcal{Q}_n + \sum_{l=0}^{n-1}\ket{\Psi_{l}}\bra{\Psi_{l}}
\: , 	
\ee
where $Q_{n} = \sum_{i \ge n} \ket{\Psi_{i}}\bra{\Psi_{i}}$ is the projector onto the orthogonal complement of the low-energy space ${\cal E}_{n}(\{ \ff S \})$, we find
\ba
\Omega^{(ij)}_{k\gamma, m\beta} 
&=&  
i \left[ 
\bra{\partial_{S_{k\gamma}}\Psi_{i}} \mathcal{Q}_n \ket{\partial_{S_{m\beta}}\Psi_{j}} 
- 
\bra{\partial_{S_{m\beta}}\Psi_{i}} \mathcal{Q}_n \ket{\partial_{S_{k\gamma}}\Psi_{j}} \right] 
\nonumber\\
&+& 
i \sum_{l=0}^{n-1} \left[ 
\bra{\partial_{S_{k\gamma}}\Psi_{i}} \ket{\Psi_{l}} \bra{\Psi_{l}} \ket{\partial_{S_{m\beta}}\Psi_{j}} 
- 
\bra{\partial_{S_{m\beta}}\Psi_{i}} \ket{\Psi_{l}} \bra{\Psi_{l}} \ket{\partial_{S_{k\gamma}}\Psi_{j}} 
\right] 
\nonumber\\
&+& 
i \sum_{l=0}^{n-1} \left[ 
\bra{\Psi_{i}} \partial_{S_{k\gamma}} \ket{\Psi_{l}} \bra{\Psi_{l}} \partial_{S_{m\beta}} \ket{\Psi_{j}} 
-  
\bra{\Psi_{i}} \partial_{S_{m\beta}} \ket{\Psi_{l}} \bra{\Psi_{l}} \partial_{S_{k\gamma}} \ket{\Psi_{j}} 
\right]
\: . 	
\ea
Noting that $\bra{\partial_{S_{m\beta}}\Psi_{i}} \ket{\Psi_{j}} = - \bra{\Psi_{i}} \partial_{S_{m\beta}} \ket{\Psi_{j}}$, we see that the last two terms on the right-hand side cancel, and thus
\be
\Omega^{(ij)}_{k\gamma, m\beta} = 
i \left[ 
\bra{\partial_{S_{k\gamma}}\Psi_{i}} \mathcal{Q}_n \ket{\partial_{S_{m\beta}}\Psi_{j}} 
- 
\bra{\partial_{S_{m\beta}}\Psi_{i}} \mathcal{Q}_n \ket{\partial_{S_{k\gamma}}\Psi_{j}} 
\right]
\: .	
\ee

\section{Numerical computation of spin-Berry curvature and connection}
\label{sec:comp}

The equations of motion \refeq{eomss} and \refeq{wfs} form a coupled, non-linear set of ordinary differential equations, which can be solved numerically by standard techniques. 
Making use of the fact that the conduction-electron system is non-interacting, however, is essential for an efficient computation of the key quantities of the electron system, namely the spin-Berry curvature and connection.

We start by specializing Eqs.\ (\ref{eq:gamma}) and (\ref{eq:sbcq}) to the single-spin case $M=1$,
\be
\langle \Omega \rangle_{\beta\gamma} 
= 
i\sum_{i,j=0}^{n-1}\alpha_{i}^{\ast}\alpha_{j}
\left(
\bra{\partial_{\beta}\Psi_{i}}\mathcal{Q}_{n}\ket{\partial_{\gamma}\Psi_{j}} - \bra{\partial_{\gamma}\Psi_{i}}\mathcal{Q}_{n}\ket{\partial_{\beta}\Psi_{j}}
\right) 
= 2 \sum_{i,j=0}^{n-1} \sum_{l \ge n} \Im{\alpha_{i}^{\ast}\alpha_{j} \bra{\Psi_{i}}\partial_{\beta}\ket{\Psi_l}\bra{\Psi_l}\ket{\partial_{\gamma}\Psi_{j}}}
\: ,
\ee
and use the identity
\be
\bra{\Psi_{i}} \partial_{\beta} \ket{\Psi_{l}}
=
\frac{\bra{\Psi_{i}}\frac{\partial \hat{H}}{\partial S_{\beta}}\ket{\Psi_{l}}}{E_{l}-E_{i}} 
\qquad (E_i \neq E_l)
\label{eq:off}
\ee
to express $\langle \boldsymbol{\Omega} \rangle$ in the form
\ba
\langle \Omega \rangle_{\beta\gamma} 
&=& 
-2 \Im \sum_{ij}\sum_{l}^{E_l \ne E_{i},E_{j}} 
\alpha_{i}^{\ast}\alpha_{j}\frac{\bra{\Psi_{i}}\frac{\partial \hat{H}}{\partial S_{\beta}}\ket{\Psi_{l}}\bra{\Psi_{l}}\frac{\partial \hat{H}}{\partial S_{\gamma}}\ket{\Psi_{j}}}{(E_{i}-E_{l})(E_{j}-E_{l})}
=
-2 \Im J^2 \sum_{ij}\sum_{l}^{E_l \ne E_{i},E_{j}}  \frac{\alpha_{i}^{\ast}\alpha_{j}\bra{\Psi_{i}}s_{i_{0}\beta}\ket{\Psi_{l}}\bra{\Psi_{l}}s_{i_{0}\gamma}\ket{\Psi_{j}}}{(E_{i}-E_{l})(E_{j}-E_{l})} 
\: .
\nonumber \\
\label{eq:gammaim}
\ea
The matrix elements can be computed by plugging in the definition of the local spin $\boldsymbol{s}_{i}=\frac{\hbar}{2}\sum_{\sigma\sigma^{\prime}}c_{i\sigma}^{\dagger}\boldsymbol{\sigma}_{\sigma\sigma^{\prime}}c_{i\sigma^{\prime}}$ and by transforming to the eigenstates of the effective hopping matrix:
\be
\label{eq: expansion of lattice c-operators in energy c-operators}
c^{\dagger}_{i\sigma} = \sum_{k\tilde{\sigma}}U^{\dagger}_{k\tilde{\sigma},i\sigma}c^{\dagger}_{k\tilde{\sigma}}
\; , \qquad 
c_{i\sigma} = \sum_{k\tilde{\sigma}}U_{i\sigma,k\tilde{\sigma}}c_{k\tilde{\sigma}}
\: .
\ee
This yields
\ba
\sum_{l}^{E_l \ne E_{i},E_{j}} 
\frac{
\bra{\Psi_{i}} s_{i_{0} \beta} \ket{\Psi_{l}} 
\bra{\Psi_{l}} s_{i_{0} \gamma} \ket{\Psi_{j}}}
{(E_{i}-E_{l})(E_{j}-E_{l})}
&=&
\frac{1}{4} \sum_{\sigma\sigma^{\prime}\tau\tau^{\prime}}{\sum}^{\prime\prime}_{k k^{\prime}q q^{\prime}\atop\tilde{\sigma}\tilde{\sigma}^{\prime}\tilde{\tau}\tilde{\tau}^{\prime}} 
U^{\dagger}_{k\tilde{\sigma},i_{0}\sigma}\sigma^{(\beta)}_{\sigma\sigma^{\prime}}
U_{i_{0}\sigma^{\prime},k^{\prime}\tilde{\sigma}^{\prime}}
U^{\dagger}_{q\tilde{\tau},i_{0}\tau}\sigma^{(\gamma)}_{\tau\tau^{\prime}}
U_{i_{0}\tau^{\prime},q^{\prime}\tilde{\tau}^{\prime}}
\times\nonumber\\
&\times&
\frac{\bra{\Psi_{i}} c^{\dagger}_{k\tilde{\sigma}}c_{k^{\prime}\tilde{\sigma}^{\prime}}c^{\dagger}_{q\tilde{\tau}}c_{q^{\prime}\tilde{\tau}^{\prime}}
\ket{\Psi_{j}}}{(E_{i}-E_{j} + \varepsilon_{q^{\prime}\tilde{\tau}^{\prime}}-\varepsilon_{q\tilde{\tau}})(\varepsilon_{q^{\prime}\tilde{\tau}^{\prime}}-\varepsilon_{q\tilde{\tau}})}
\: ,
\ea
where $\sum^{\prime\prime}$ means that the indices $k, k^{\prime},q, q^{\prime}, \tilde{\sigma}, \tilde{\sigma}^{\prime}, \tilde{\tau}, \tilde{\tau}^{\prime}$ can only take values such that $c^{\dagger}_{k^{\prime}\tilde{\sigma}^{\prime}}c_{k\tilde{\sigma}}\ket{i}$ and $c^{\dagger}_{q\tilde{\tau}}c_{q^{\prime}\tilde{\tau}^{\prime}}\ket{j}$ are not contained in the low-energy subspace.
For the summation indices it is required that
\be
(k,\tilde{\sigma}) \neq (k^{\prime},\tilde{\sigma}^{\prime}) 
\quad \mbox{and} \quad 
(q,\tilde{\tau}) \neq (q^{\prime},\tilde{\tau}^{\prime})
\ee
since $i\neq l$ and $j\neq l$. 
Plugging this into \eqref{eq:gammaim} gives an expression that can be evaluated straightforwardly by numerical means.

We also have to compute the Berry connection, i.e., the matrix elements $\bra{\Psi_{i}} \partial_\beta \ket{\Psi_{j}}$ in \refeq{wfs} (see also \refeq{con}).
For $i \neq j$ we can again use \eqref{eq:off}, since the single-particle energies are generically nondegenerate for finite $J$ and since this implies that states $\ket{\Psi_{i}}$ and $\ket{\Psi_{j}}$ with $E_i = E_j$ must differ in more than one single-particle eigenstate.
For $i=j$, one the other hand, $\bra{\Psi_{i}} \partial_\beta \ket{\Psi_{i}}$ mast be computed differently.
We exploit that the many-particle state $\ket{\Psi_{i}}$ is a Slater determinant:
\be
\ket{\Psi_{i}} = c^{\dagger}_{n_{1}}c^{\dagger}_{n_{2}} \cdots c^{\dagger}_{n_{N}} \ket{\text{vac}} 
\: .
\ee
Therewith, we get
\be
\partial_{S_{\beta}}\ket{\Psi_{i}} 
= 
\sum_{i=1}^{N} c^{\dagger}_{n_{1}} \cdots (\partial_{S_{\beta}}c^{\dagger}_{n_{i}}) \cdots c^{\dagger}_{n_{N}} \ket{\text{vac}}
\labeq{cder}
\ee
with
\begin{align}
\partial_{S_{\beta}}c^{\dagger}_{n_{i}} 
= 
\partial_{S_{\beta}} \sum_{j\sigma} U_{j\sigma,n_{i}} c^{\dagger}_{j\sigma} 
= \sum_{j\sigma} (\partial_{S_{\beta}}U_{j\sigma,n_{i}}) c^{\dagger}_{j\sigma} 
= \sum_{j\sigma}\sum_{m} (\partial_{S_{\beta}}U_{j\sigma,n_{i}}) U^{\dagger}_{m,j\sigma} c^{\dagger}_{m} 
= \sum_{m} (U^{\dagger}\partial_{S_{\beta}}U)_{mn_{i}} c^{\dagger}_{m}
\: .
\end{align}	
Multiplying \refeq{cder} with $\bra{\Psi_{i}}$ from the left yields
\ba
\bra{\Psi_{i}} \partial_\beta \ket{\Psi_{i}} 
&=& 
\sum_{i=1}^{N}\sum_{m} (U^{\dagger}\partial_{S_{\beta}}U)_{mn_{i}} \underbrace{\bra{\text{vac}}c_{n_{N}} \cdots c_{n_{i}} \cdots c_{n_{1}}c^{\dagger}_{n_{1}} \cdots c^{\dagger}_{m} \cdots c^{\dagger}_{n_{N}}\ket{\text{vac}}}_{\delta_{n_{i}m}} 
\nonumber\\
&=&
\sum_{i=1}^{N} (U^{\dagger}\partial_{S_{\beta}}U)_{n_{i}n_{i}}
= {\sum_{n}}^{\prime} (U^{\dagger}\partial_{S_{\beta}}U)_{nn} 
= \sum_{n} (U^{\dagger}\partial_{S_{\beta}}U)_{nn} \bra{\Psi_{i}}\hat{n}_{n}\ket{\Psi_{i}}
\: ,
\ea
where $\sum_{n}^{\prime}$ indicates that the sum only contains those single-particle states that are occupied in the many-particle state $\ket{\Psi_{i}}$. 
The derivative of the $U$-matrix can be computed by standard numerical means.

\end{widetext}

\section{Time-reversal symmetric ground state}
\label{sec:trs}

We consider the minimal model with Hamiltonian $H$ given by \refeq{min}. 
For $J=0$ the (electronic part of the) model is invariant under SU(2) spin rotations. 
For a given direction of the classical spin, say $\ff S = S \ff e_{z}$, and for $J>0$ the symmetry is breaks down to a U(1) symmetry under spin rotations around the $z$ axis.
As argued in the main text, the local spin-dependent perturbation is not strong enough to spin-polarize the system, irrespective of the coupling strength $J$. 
In this case the ground state of $H$ is invariant under time reversal, as is shown in the following:

The antiunitary operator $\Theta$ representing time reversal in Fock space is defined via its action on the creation and annihilation operators as 
\be
\Theta c^\dagger_{i\uparrow} \Theta^\dagger = c^\dagger_{i\downarrow} \; , \quad
\Theta c^\dagger_{i\downarrow} \Theta^\dagger = -c^\dagger_{i\uparrow} 
\: , 
\ee
where $i$ refers to lattice sites and $\sigma=\uparrow, \downarrow$ to the spin projection with respect to the $z$ axis.
Due to the remaining U(1) symmetry, the Hamiltonian can be diagonalized in the spin-$\uparrow$ and spin-$\downarrow$ sectors  separately, i.e., the single-particle eigenstates $c^{\dagger}_{k\sigma} | \mbox{vac} \rangle$ of $H$ are obtained via a spin-diagonal and spin-independent unitary transformation:
\be
  c^{\dagger}_{k\sigma} = \sum_{i} U_{ik} c^{\dagger}_{i\sigma} \: . 
\ee
For the model \refeq{min} with $\ff S = S \ff e_{z}$, the effective hopping matrix \refeq{teff} is real and symmetric, and we can thus assume a real and orthogonal transformation matrix $U$. 
The creation operators referring to the eigenbasis of $H$ in the one-particle subspace thus transform as 
\be
\Theta c^\dagger_{k\uparrow} \Theta^\dagger = c^\dagger_{k\downarrow} 
\; , \quad 
\Theta c^\dagger_{k\downarrow} \Theta^\dagger = -c^\dagger_{k\uparrow} 
\: 
\ee
under time reversal.

For even $N$, the ground state of $H$ is the Slater determinant
\be
\ket{\Psi_{0}} = \prod^{\rm occ.}_k c^\dagger_{k\uparrow} \prod^{\rm occ.}_{k^\prime} c^\dagger_{k^\prime\downarrow} \ket{\text{vac}} \: , 
\ee
where $\ket{\text{vac}}$ is the time-reversal invariant vacuum, $k=1,...,N_{\uparrow}$, and $k'=1,..., N_{\downarrow}$ with $N_{\uparrow} = N_{\downarrow} = N/2$, as the ground state is unpolarized.
Applying $\Theta$ yields
\be
\Theta \ket{\Psi_{0}} = (-1)^{N_{\downarrow}} \prod^{\rm occ.}_k c^\dagger_{k\downarrow} \prod^{\rm occ.}_{k^\prime} c^\dagger_{k^\prime\uparrow} \ket{\text{vac}} \: , 
\ee
and, after reordering, 
\be
\Theta \ket{\Psi_{0}} = (-1)^{N_{\uparrow} N_{\downarrow}} (-1)^{N_{\downarrow}} \prod^{\rm occ.}_{k^\prime} c^\dagger_{k^\prime\uparrow} 
\prod^{\rm occ.}_k c^\dagger_{k\downarrow} \ket{\text{vac}} \: .
\ee
For $N_{\uparrow} = N_{\downarrow} = N/2$, however, the total sign is $+1$, and hence the ground state is time-reversal symmetric
\be
\Theta \ket{\Psi_{0}} =  \ket{\Psi_{0}} \: .
\ee

\end{document}